\documentclass[10pt,
superscriptaddress,
twocolumn,
amsmath,amssymb,
aps,
pra,
]{revtex4-2}

\usepackage{graphicx,framed} % Include figure files
\usepackage{dcolumn} % Align table columns on decimal point
\usepackage{bm} % bold math
\usepackage{braket}
\usepackage{color}
\usepackage{amsthm}
\usepackage{hyperref} % add hypertext capabilities
\usepackage[normalem]{ulem} %to strike the words

\usepackage{amsthm}

\usepackage{comment}

\usepackage{float}
\makeatletter
\let\newfloat\newfloat@ltx
\makeatother
\usepackage{algorithm}
\usepackage[noend]{algpseudocode}
\algdef{SE}[DOWHILE]{Do}{doWhile}{\algorithmicdo}[1]{\algorithmicwhile\ #1}
\makeatletter
\def\algbackskip{\hskip-\ALG@thistlm}
\makeatother

\definecolor{lightblue}{RGB}{73,151,208}
\definecolor{crimson}{RGB}{140,41,53}

\hypersetup{
    colorlinks,
    linkcolor={crimson},
    citecolor={lightblue},
    urlcolor={lightblue}
}

\begin{document}

\preprint{}

\title{Unpolarized prethermal discrete time crystal}

\author{Takeru Yokota}
\email{takeru.yokota@oit.ac.jp}
\affiliation{RIKEN Center for Quantum Computing, Wako, Saitama 351-0198, Japan}
\affiliation{Faculty of Engineering, Osaka Institute of Technology, Asahi-ku, Osaka, 535-8585, Japan}
\affiliation{RIKEN Center for Interdisciplinary Theoretical and Mathematical Sciences, Wako, Saitama, 351-0198, Japan}

\author{Tatsuhiko N. Ikeda}
\email{tatsuhiko\_ikeda@zen.ac.jp}
\affiliation{RIKEN Center for Quantum Computing, Wako, Saitama 351-0198, Japan}
\affiliation{Faculty of Social Informatics, ZEN University, Zushi, Kanagawa, 249-0007, Japan}
\affiliation{Department of Applied Physics, Hokkaido University, Sapporo, Hokkaido, 060-8628, Japan}

\date{\today}%

\begin{abstract}
Prethermal discrete time crystals (DTCs) are a novel phase of periodically driven matter that exhibits robust subharmonic oscillations without requiring disorder. However, previous realizations of prethermal DTCs have relied on the presence of polarization, either spontaneous or induced. Here, we introduce a new class of prethermal DTCs termed ``unpolarized prethermal discrete time crystals'' (UPDTCs) that arise without any uniform/staggered polarization and propose an experiment to observe them in current trapped-ion quantum simulators. By studying a model of trapped ions using a quantum circuit simulator, we demonstrate that robust period-doubled dynamics can persist in the autocorrelation function of the staggered magnetization, even though its expectation value does not exhibit such dynamics. The period-doubled dynamics is not explained by the classical picture of flipping spins but by quantum fluctuations. We establish that UPDTCs are exponentially long-lived in the high-frequency driving regime, a hallmark of prethermalization. These results expand the known phenomenology and mechanism of prethermal time crystals and underscore the role of quantum effects in stabilizing novel nonequilibrium phases.
\end{abstract}

\maketitle

\section{Introduction}
% \lsec{Introduction}
Time crystals have attracted tremendous interest in recent years as a novel form of nonequilibrium quantum matter that spontaneously breaks time translation symmetry~\cite{Wilczek2012,Bruno2013,Watanabe2015,Sacha2017,Khemani2019}. The fundamental concept of time crystals, originally proposed by Wilczek~\cite{Wilczek2012}, has captivated researchers from fields as varied as condensed matter physics~\cite{Watanabe2015,Bruno2013}, atomic and molecular physics~\cite{Zhang2017,Machado2020,Kyprianidis2021}, quantum information science~\cite{Ippoliti2021,Xiang2024}, and statistical mechanics~\cite{Tindall2020,Gambetta2019,Zhao2019}. In periodically driven or Floquet systems~\cite{Goldman2014,Holthaus2015,Bukov2015,Eckardt2017,Oka2019}, discrete time crystals (DTCs) can emerge, characterized by observables that oscillate at a longer period than the driving period~\cite{Else2016,Yao2017,Else2017}. However, isolated driven many-body systems generically heat up to infinite temperature~\cite{DAlessio2014,Lazarides2014,Kim2014}, destroying the DTC phase. To stabilize DTCs, robust mechanisms like many-body localization (MBL)~\cite{Yao2017,Zhang2017} and Floquet prethermalization~\cite{Else2017,Machado2020,Kyprianidis2021} have been employed to prevent thermalization on experimentally relevant timescales (see also, e.g., Refs.~\cite{Holthaus1994,Mizuta2018,Tindall2020,Chinzei2020,Gambetta2019,Yu2019,Zhao2019,Lledo2020,O’Sullivan2020,Kuros2020,Pizzi2020,Basu2022,Liu2023,Xie2024a,Xie2024b,Autti2018,Autti2021,Autti2022,Collura2022,Gargiulo2024} for other mechanisms and experiments).

Prethermal discrete time crystals (PDTCs) have emerged as a particularly promising platform for realizing long-lived time crystalline order without requiring localization. In a PDTC, periodic driving is applied at high frequency such that the system prethermalizes to a metastable state instead of fully thermalizing. Two main mechanisms for PDTCs have been established. One relies on spontaneous symmetry breaking (SSB) in the prethermal Hamiltonian, leading to robust period doubling for initial states that break a $\mathbb{Z}_2$ symmetry \cite{Else2017}. The other leverages prethermalization without symmetry breaking, stabilizing DTC order for U(1) symmetric initial states under strong longitudinal fields that enforce longitudinal magnetization \cite{Luitz2020,Stasiuk2023}.

However, these PDTC phases proposed thus far have both relied on an oscillating polarization (uniform or staggered) to diagnose the time crystalline behavior. The underlying physics can be intuitively understood in terms of precession of classical spin vectors~\cite{Pizzi2021,Pizzi2021b,Ye2021,Howell2019}. An open question remains whether PDTCs can exist without polarization, i.e. for quantum paramagnetic states that do not break any symmetries and have no net magnetization. Realizing such phases would reveal fundamentally new mechanisms of prethermal time crystals.

In this paper,
% In this Letter, 
we predict the existence of a new class of PDTCs without any uniform/staggered polarization, which we term unpolarized prethermal discrete time crystals (UPDTCs). We show that robust time crystalline signatures can emerge in the autocorrelation function of the staggered magnetization for paramagnetic initial states governed by a transverse-field Ising model with long-range interactions for trapped-ion quantum simulators~\cite{Kyprianidis2021}. Remarkably, this occurs even when the expectation value of the uniform/staggered magnetization strictly vanishes. The key insight is that quantum fluctuations in the magnetization can reveal DTC-like dynamics even when its expectation value vanishes. Our results establish UPDTCs as a qualitatively distinct incarnation of prethermal time crystals, arising from genuine quantum effects without a classical counterpart.

This paper is organized as follows: In Sec.~\ref{sec: problem}, we describe the problem setting. The results for UPDTC signals in autocorrelation are presented in Sec.~\ref{sec: UPDTC}. The origin and interpretation of the obtained signals are discussed in Sec.~\ref{sec: origin}. In Sec.~\ref{sec: experiment}, we describe the experimental setup for observing UPDTC. Section \ref{sec: conclusions} is devoted to the conclusions.

\begin{figure*}
    \centering
    \includegraphics[width=1.0\linewidth]{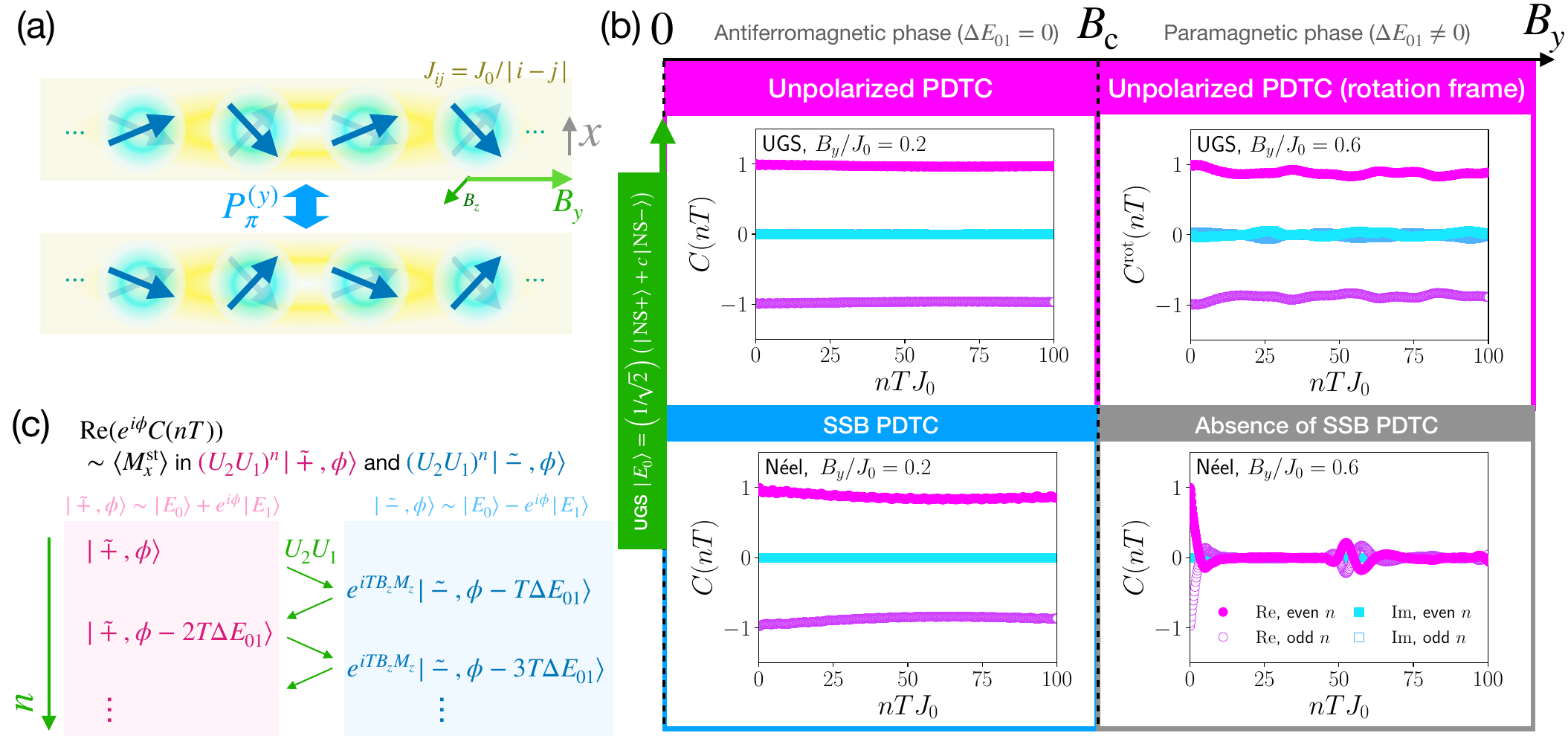}
    \caption{Setup and main results. (a) Illustration of our model. The direction of spin fluctuations along the \(x\)-axis changes with \(\pi\)-pulses, as captured by the autocorrelation function \(C(nT)\). (b) Results of \(C(nT)\) for the UGS $\Ket{E_0}$ and N\'eel state \(\Ket{\mathrm{NS}+}\) at \(B_y/J_0 = 0.2\) (antiferromagnetic phase) and \(B_y/J_0 = 0.6\) (paramagnetic phase) with \(TJ_0 = 0.1\), along with the corresponding signal types. (c) Illustration of the time evolution of \(\Ket{\tilde{\pm},\phi}\), whose staggered magnetization determines \(C(nT)\). Here we used $U_1\approx e^{-iTH_\mathrm{eff}^{(0)}}e^{-iTB_z M_z}$ with \(M_z\equiv\sum_{i=1}^L\sigma^z_i\).}
    \label{fig:main-results}
\end{figure*}
\section{Problem setting \label{sec: problem}}
% \lsec{Problem setting}
Inspired by the setups implemented in ion trap experiments \cite{Kyprianidis2021}, we consider an \( L \)-site spin-1/2 chain with long-range interactions periodically driven by \( \pi \)-pulses around the \( y \)-axis with period \( T \). The time evolution of a state \(\Ket{\psi}\) over a duration \(nT\) (\(n \in \mathbb{Z}_{\geq 0}\)) is described by $U(nT)\Ket{\psi}$, where $U(nT)\equiv (U_2U_1)^n$ and
\begin{align}
    U_1&=e^{-iT\left(\sum_{i<j}^{L}J_{ij}\sigma^x_i \sigma^x_j + B_y\sum_{i=1}^L\sigma^y_i + B_z\sum_{i=1}^L\sigma^z_i\right)},
    \\
    U_2&=P_\pi^{(y)}=e^{-i \frac{\pi}{2}\sum_{i=1}^L\sigma^y_i}.
\end{align}
Here, \( B_y \) and \( B_z \) represent the applied transverse magnetic fields, and the interactions are long-range antiferromagnetic \( J_{ij} = J_0 / |i-j| \) (\( J_0 > 0 \)) as in the experiments~\cite{Kyprianidis2021}. We assume that $U_2$ is applied instantaneously and the physical time during $U_2 U_1$ is $T$, so we call $T$ and $2\pi/T$ the driving period and frequency, respectively. We fix \( L = 21 \) and \( B_z / J_0 = 1 \) unless stated otherwise. Our setup is illustrated in Fig.~\ref{fig:main-results}(a).

Before studying DTC-like behaviors, we discuss its approximate effective dynamics when $T$ is small. To average out the effect of $U_2$, we consider the effective Hamiltonian over the two periods, \(H_{\mathrm{eff}} = \frac{i}{2T} \ln (U_2 U_1 U_2 U_1)\).
At the lowest order of the series expansion for $T$ (i.e., the high-frequency expansion~\cite{Bukov2015}), \(H_{\mathrm{eff}}\) corresponds to the transverse-field Ising model:
\begin{align}
    H_{\mathrm{eff}}=\sum_{i<j}^{L}J_{ij}\sigma^x_i \sigma^x_j + B_y\sum_{i=1}^L\sigma^y_i + O(T).
\end{align}
To represent the leading terms, we introduce the notation \( H_{\mathrm{eff}}^{(0)} = \sum_{i<j}^{L} J_{ij} \sigma^x_i \sigma^x_j + B_y \sum_{i=1}^L \sigma^y_i \), and denote its eigenenergies and eigenstates as $H_{\mathrm{eff}}^{(0)}\ket{E_k}=E_k\ket{E_k}$ $(k=0,1,\dots)$. 
The (leading-order) effective Hamiltonian \(H_{\mathrm{eff}}^{(0)}\) is known~\cite{Koffel2012} to undergo an antiferromagnetic transition in the $L\to\infty$ limit at \(B_y = B_{\mathrm{c}}\), which corresponds to \(B_{\mathrm{c}} / J_0 = 0.52\pm0.01\) as shown in Appendix \ref{app: transition point}. Here, the order parameter is the staggered magnetization
\begin{align}
M_x^{\mathrm{st}} = \sum_{i=1}^L (-1)^{i-1}\sigma_i^x,
\end{align}
instead of the uniform magnetization \(M_x = \sum_{i=1}^L \sigma_i^x\) in ferromagnetic systems.
For $B_y>B_c$, the ground state is paramagnetic and symmetric, whereas the $\mathbb{Z}_2$ symmetry, \([P_\pi^{(y)}, H_{\mathrm{eff}}^{(0)}] = 0\), is broken for $B_y<B_c$, and there are a pair of degenerate N\'{e}el-like ground states. At finite $L$, however, the ground state $\ket{E_0}$ is always symmetric and unpolarized, i.e., 
\begin{align}
\bra{E_0}M_x^\mathrm{st}\ket{E_0}=0.    
\end{align}
We call this symmetric $\Ket{E_0}$ the unpolarized ground state (UGS).

Within the two periods, this model expectedly accommodates an SSB PDTC for $B_y<B_c$. Let us imagine we prepare one of the N\'{e}el-like states and let it evolve under alternate unitaries $U_1$ and $U_2$. The state is nearly unchanged through $U_1$, whereas $U_2$ transforms it into the other N\'{e}el-like state. Consequently, alternate sign changes appear in the nonzero expectation value \(\Braket{M_x^{\mathrm{st}}}\neq 0\) for the staggered magnetization \(M_x^{\mathrm{st}} \) at the doubled period $2T$.
This intuition might lead us to anticipate that, either without SSB ($B_y>B_c$) or without the use of symmetry-broken states, DTC-like signals would disappear. This is indeed true as long as we only observe the staggered magnetization. However, we will show below that DTC-like signals are hidden in its autocorrelation function even without SSB or symmetry-broken states. 

\section{UPDTC in autocorrelation \label{sec: UPDTC}}
% \lsec{UPDTC in autocorrelation}
To uncover the DTC-like signal without polarization, we look into the (normalized) stroboscopic autocorrelation function of \(M_x^{\mathrm{st}}\),
\begin{align}
    C(nT)=\frac{1}{\mathcal{N}^{2}}\Braket{M_x^{\mathrm{st}}(nT)M_x^{\mathrm{st}}}_\psi,
\end{align}
where \(M_x^{\mathrm{st}}(nT) = U(nT)^{\dagger} M_x^{\mathrm{st}} U(nT)\), \(\mathcal{N}=C(0)^{1/2}=\sqrt{\Braket{(M_x^{\mathrm{st}})^2}_\psi}\), and $\langle \cdots \rangle_\psi$ denotes the expectation value for the inital state $\Ket{\psi}$ (see Appendix \ref{app: method time evolution} for methods). We note that autocorrelation has been used as a diagnostic of time crystallinity in the literature~\cite{Watanabe2015,Luitz2020,Kyprianidis2021}.
Also, we will show how to measure this quantity later in this paper. 
% in this Letter. 
For comparison, we also perform the computation for the N\'eel state, which is expected to exhibit SSB PDTC. We use \(\Ket{\mathrm{NS}+} = \Ket{+}_1 \otimes \Ket{-}_2 \otimes \Ket{+}_3 \cdots\), where \(\Ket{+}_i\) (\(\Ket{-}_i\)) represents the spin at site \(i\) pointing in the positive (negative) \(x\)-direction, as the N\'eel state. The other N\'eel state, \(\Ket{\mathrm{NS}-} = \Ket{-}_1 \otimes \Ket{+}_2 \otimes \Ket{-}_3 \cdots\), behaves similarly.

Our main result, carefully identified below, is stated as
\begin{align}\label{eq:main}
    C(nT) \approx (-1)^n e^{-in \Omega T} f(n)\qquad \text{for $\ket{\psi}=\ket{E_0}$},
\end{align}
where $\Omega$ is real and $f(n)$ is a slowly decaying nonnegative function with $f(0)=1$.
Since the sinusoidal factor $e^{-in\Omega T}$ can be eliminated in an appropriate rotating frame
\begin{align}\label{eq:main_rot}
    C^{\mathrm{rot}}(nT) \equiv e^{i\Omega nT}C(nT) \approx (-1)^n f(n)
\end{align}
that coincides with the DTC signal, we call the behavior~\eqref{eq:main} to be DTC-like.
Remarkably, Eq.~\eqref{eq:main} holds even in the paramagnetic phase, where the N\'{e}el state no longer shows DTC-like behaviors.
We emphasize again that the UGS is not polarized, and the DTC-like behavior~\eqref{eq:main} is not derived from previously known mechanisms including classical spins~\cite{Pizzi2021,Pizzi2021b,Ye2021}.

\subsection{Early-time dynamics}
% \lsec{Early-time dynamics}
We first focus on the early stages of evolution and verify the main claim~\eqref{eq:main}. There $f(n)\approx 1$, and we expect $C(nT)\approx (-1)^n e^{-in\Omega T}$.
In the antiferromagnetic phase, the autocorrelation $C(nT)$ exhibits DTC behaviors for both initial states, the UGS and a N\'{e}el state.
The left panels of Fig.~\ref{fig:main-results}(b) show \(C(nT)\) for each state at \(B_y/J_0 = 0.2<B_c/J_0\) and \(TJ_0 = 0.1\). As a magnetically ordered state is favored as the low-energy state in the antiferromagnetic phase, the N\'eel state exhibits SSB PDTC. Meanwhile, although the UGS is defined as the ground state invariant under \(P_\pi^{(y)}\), it also displays DTC signals through \(C(nT)\), as shown in the upper-left panel, meaning the emergence of the UPDTC with $\Omega=0$.

\begin{figure}
    \centering
    \includegraphics[width=1.0\linewidth]{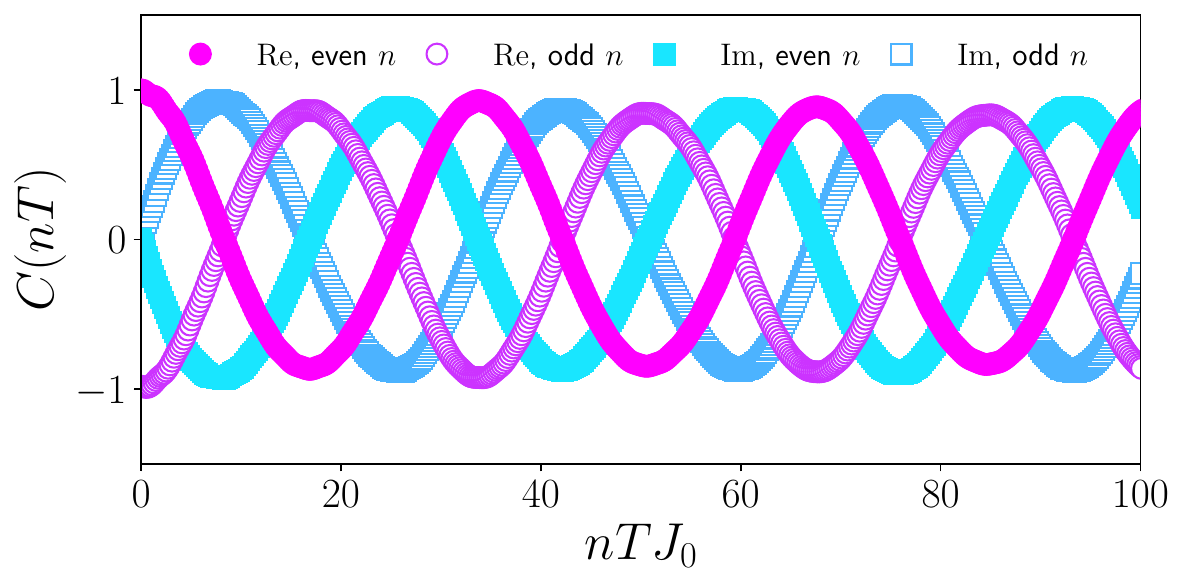}
    \caption{Real and imaginary parts of \(C(nT)\) of the UGS at \(TJ_0 = 0.1\) and \(B_y/J_0 = 0.6\).}
    \label{fig:C_ri}
\end{figure}

In the paramagnetic phase, in contrast, only the UGS shows DTC-like signals with $\Omega\neq0$ while N\'{e}el states lose such signals as expected.  Figure \ref{fig:C_ri} shows the results for \(C(nT)\) of the UGS at \(TJ_0 = 0.1\) and \(B_y/J_0 = 0.6>B_c/J_0\). On top of the sign-flip behavior $(-1)^n$, we observe slower sinusoidal envelopes given by $e^{-in\Omega T}$. If so, in the rotating frame with a well-chosen frequency $\Omega$, the DTC behavior $(-1)^n$ can be extracted. This is indeed true as shown in the upper-right panel of Fig.~\ref{fig:main-results}(b). There, \(\Omega\) is determined as the average phase change per step in \(C(nT)\), calculated as \(\Omega = -\frac{1}{TN_{\mathrm{t}}} \sum_{n=0}^{N_{\mathrm{t}}-1} \mathrm{arg}\left(-\frac{C((n+1)T)}{C(nT)}\right)\), with the total number of time-evolution steps \(N_{\mathrm{t}}\) (we will theoretically identify $\Omega$ below). This demonstrates that the UPDTC remains a stable signal in both the antiferromagnetic and paramagnetic phases, provided a phase-rotating frame is introduced. This contrasts with the N\'eel state, whose \(C(nT)\) at \(B_y/J_0 = 0.6\) and \(TJ_0 = 0.1\), shown in the lower-right panel of Fig.~\ref{fig:main-results}(b), exhibits rapid initial signal decay. While signals are detected even after the initial decay, they are short-lived and lack the order characteristic of DTCs. We note that similar results are obtained for short-range interactions (see Appendix \ref{app: short range}).

Notably, the initial signal's decay of the N\'eel state has also been observed experimentally~\cite{Kyprianidis2021}, where the measured quantity aligns with \(C(nT)\) of the N\'eel state. The experimental setup resembles our model with \(B_z/J_0 = 0.6\) and \(B_y/J_0 = 1.5\), where the effective Hamiltonian ground state is expected to exhibit paramagnetism. Our finding asserts that, even in such a paramagnetic regime, the DTC-like signal could be uncovered if the UGS is prepared and the autocorrelation is measured.

%%%%%%%%%%%%%%%%%%%%%%%%%%%%%%%%%%%%%%%%%%%%%%%%%
%%%%%%%%%%%%%%%%%%%%%%%%%%%%%%%%%%%%%%%%%%%%%%%%%
%%%%%%%%%%%%%%%%%%%%%%%%%%%%%%%%%%%%%%%%%%%%%%%%%
\begin{figure}
    \centering
    \includegraphics[width=1.0\linewidth]{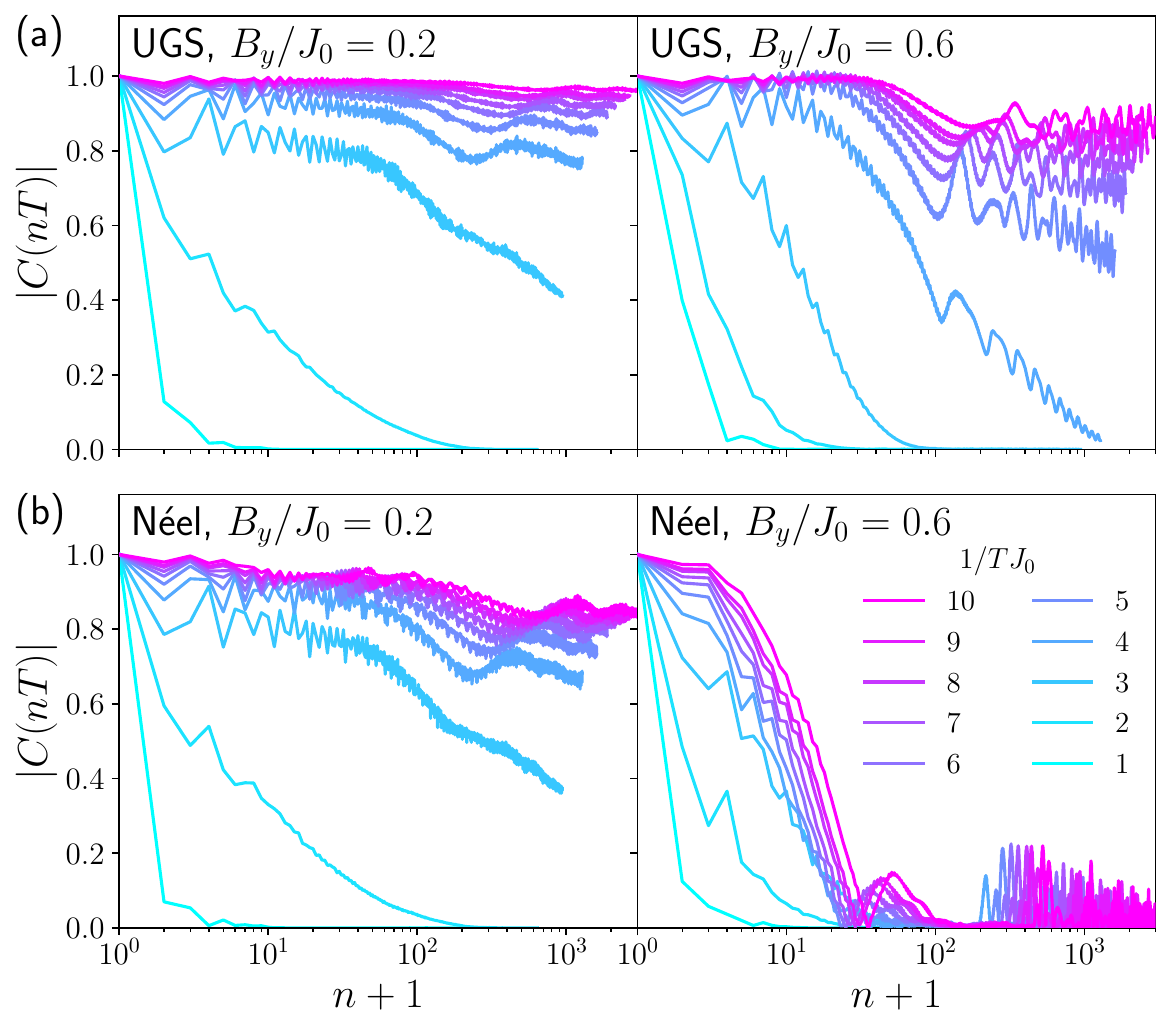}
    \caption{
    Time profile of \(|C(nT)|\) at various driving frequencies \(1/TJ_0\) for (a) the UGS and (b) the N\'eel state \(\Ket{\mathrm{NS}+}\) in \(0\leq n \leq 300/TJ_0\). For each state, results are shown for \(B_y/J_0 = 0.2\) (antiferromagnetic phase) and \(B_y/J_0 = 0.6\) (paramagnetic phase).}
    \label{fig:absC}
\end{figure}

\subsection{Late-time dynamics}
% \lsec{Late-time dynamics}
Let us now focus on $f(n)\approx |C(nT)|$ in Eq.~\eqref{eq:main} represending the late-time dynamics. Specifically, we examine whether the signal exhibited by the UGS follows Floquet prethermalization, where increasing the driving frequency prevents thermalization to infinite temperature due to periodic driving, thereby exponentially extending the signal's lifetime \cite{Abanin2015,Mori2016,Kuwahara2016,Abanin2017,Avdoshkin2020}. Studies of the kicked Ising model \cite{Ikeda2024} suggest that, for sufficiently short driving periods, the ground state of the effective Hamiltonian is robust against thermalization. According to this mechanism, Floquet prethermalization is expected to occur for the UGS, regardless of the phase exhibited by the effective Hamiltonian. From Fig.~\ref{fig:absC}(a), it is evident that the exponential prolongation of the signal lifetime for increasing \(1/T\) is realized not only in the antiferromagnetic phase but also in the paramagnetic phase. This contrasts with the N\'eel state case shown in Fig.~\ref{fig:absC}(b), where the increase in the initial signal's lifetime ceases immediately upon entering the paramagnetic phase, resulting in the absence of the SSB PDTC observed in Fig.~\ref{fig:main-results}(b) (see 
Appendix \ref{app: lifetime}
% Ref.~\cite{sm} 
for a quantitative verification). Also, we confirmed that the long-lived DTC-like signal survives even in the presence of small disorder in \(B_{y,z}\), as shown in Appendix \ref{app: robustness}.
% \cite{sm}.

\begin{figure}
    \centering
    \includegraphics[width=1.0\linewidth]{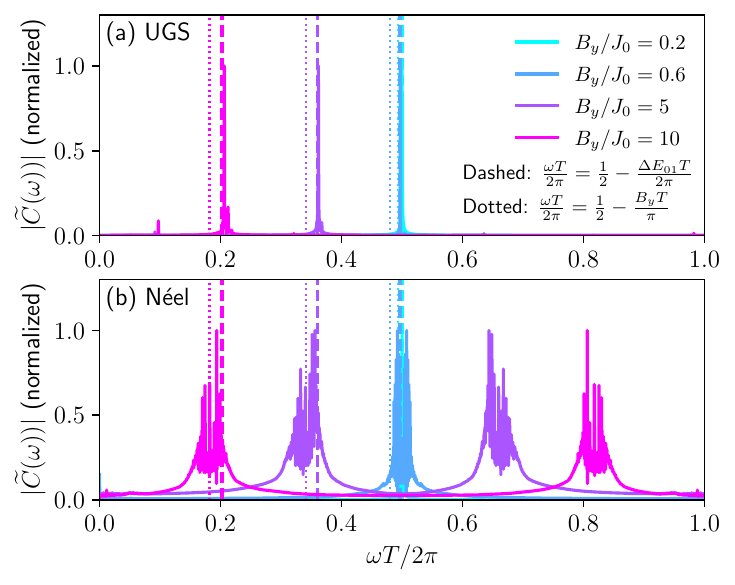}
    \caption{The magnitude spectra \(|\widetilde{C}(\omega)|\) for various \(B_y\) and \(TJ_0=0.1\) in (a) the UGS and (b) the N\'eel state \(\Ket{\mathrm{NS}+}\). The signal for each \(B_y\) is normalized by the maximum value \(\max_{\omega}|\widetilde{C}(\omega)|\). 
    The results are calculated from \(C(nT)\) within the range \(0 \leq n \leq 3200\). Vertical dashed (dotted) lines are drawn at \(\omega T / 2\pi = 1/2 - T\Delta E_{01} / 2\pi\) (\(\omega T / 2\pi = 1/2 - B_y T / \pi\)) for each \(B_y\). Solid, dashed, and dotted lines of the same color correspond to results for the same \(B_y\).}
    \label{fig:fourier}
\end{figure}

\subsection{Sinusoidal oscillation}
% \lsec{Sinusoidal oscillation}
Let us finally investigate the remaining piece $e^{-in\Omega T}$ in our main result~\eqref{eq:main}. Since $\Omega=0$ in the antiferromagnetic phase, we examine how $\Omega$ varies when $B_y$ increases to go deeper in the paramagnetic phase.
An important observation is 
\begin{align}\label{eq:Omega}
    \Omega \approx \Delta E_{01}.
\end{align}
In Fig.~\ref{fig:C_ri}, $\Omega$ is given by \(\Omega/J_0 = 0.186\), which is close to \(\Delta E_{01}/J_0 \equiv (E_1-E_0)/J_0  = 0.210\).
To compare \(\Omega\) and \(\Delta E_{01}\) for various \(B_y\), the matnitude spectrum \(|\widetilde{C}(\omega)|\) for the autocorrelation \(C(nT)\) is shown in Fig.~\ref{fig:fourier}(a). The spectrum has a single sharp peak in all cases of \(B_y\). For \(B_y/J_0 = 0.2\) (antiferromagnetic phase), a peak appears at \(\omega = \pi/T\), while for \(B_y/J_0 \gtrsim 0.52\) (paramagnetic phase), the peak position shifts away from \(\omega = \pi/T\), with the shift giving rise to the phase rotation frequency \(\Omega\). Regardless of \(B_y\), the peak position closely matches \(\omega = \pi/T - \Delta E_{01}\), as indicated by the dashed vertical line, confirming Eq.~\eqref{eq:Omega}.
Note that the peak position does not coincide with the position \(\omega = \pi/T - 2B_y\) indicated by the dotted vertical line. This suggests that the frequency $\Omega$ originates not from the Rabi oscillations seen in a non-interacting system \cite{Zhang2017} but from the many-body effect \(\Delta E_{01}\). Additionally, while there is a slight difference between the peak position and \(\omega = \pi/T - \Delta E_{01}\), this is attributed to the use of an approximate effective Hamiltonian, and as \(T\) becomes smaller, this discrepancy decreases, as shown in Appendix \ref{app: peak}.
% \cite{sm}.

The single peak structure in \(|\widetilde{C}(\omega)|\) for the UGS is in contrast to the N\'eel state. Figure \ref{fig:fourier}(b) plots \(|\widetilde{C}(\omega)|\) for the N\'eel state with the same choices of \(B_y\) as for the UGS. In the antiferromagnetic phase (\(B_y/J_0 = 0.2\)), the N\'eel state, like the UGS, shows a single peak at \(\omega = \pi/T\), which represents the SSB PDTC. However, in the paramagnetic phase (\(B_y/J_0 \gtrsim 0.52\)), in contrast to the UGS, the N\'eel state exhibits a more complex structure with multiple peaks around \(\omega = \pi/T \pm \Delta E_{01}\) and \(\omega = \pi/T \pm 2B_y\). This indicates that, in the paramagnetic phase, the N\'eel state is no longer a low-energy state and includes many excited states. Note that the N\'eel state's \(|\widetilde{C}(\omega)|\) exhibits a symmetric structure around \(\omega = \pi/T\) because \(C(nT)\) is real, unlike the UGS.

\section{Origin and interpretation of the DTC-like signal \label{sec: origin}}
% \lsec{Origin and interpretation of the DTC-like signal}
We theoretically uncover the origin of the DTC-like behavior~\eqref{eq:main} without polarization. When \(T\) is sufficiently small, the time evolution of the system is well approximated by \(e^{-i H_{\mathrm{eff}}^{(0)}nT}\) before Floquet heating matters~\cite{Kuwahara2016,Abanin2017}. Within this approximation, as derived in Appendix \ref{app: derivation}, \(C(nT)\) can be expressed as
% ~\cite{sm}
\begin{align}
    \label{eq:Cspectral}
    C(nT)\approx &(-1)^n\sum_{m\geq 1}e^{-inT\Delta E_{0m}} 
    \left|\Braket{E_m|E_0'}\right|^2,
\end{align}
where \(\Delta E_{0m}\equiv E_m-E_0\) and
\begin{align}
    \Ket{E_0'}=\frac{M^{\mathrm{st}}_x\Ket{E_0}}{\|M^{\mathrm{st}}_x\Ket{E_0}\|}.
\end{align}
Comparing Eq.~\eqref{eq:Cspectral} with our observation of Eqs.~\eqref{eq:main} and \eqref{eq:Omega}, we anticipate that
\begin{align}
    \label{eq:E1approx}
    \Ket{E_0'}
    \approx
    \Ket{E_1}
\end{align}
is the underlying mechanism.
This relation holds exactly in the case of \(B_y = 0\). For \(B_y = 0\), we can express \(\Ket{E_0} = (1/\sqrt{2})(\Ket{\mathrm{NS}+} + c\Ket{\mathrm{NS}-})\) and \(\Ket{E_1} = (1/\sqrt{2})(\Ket{\mathrm{NS}+} - c\Ket{\mathrm{NS}-})\), where \(c = 1\) (\(c = i\)) for even (odd) \(L\), making Eq.~\eqref{eq:E1approx} exact. For \(B_y = \infty\), the perturbative analysis gives
% ~\cite{sm}
\begin{align}
    |\Braket{E_1|E_0'}|^2=\frac{8}{\pi^2}+O(L^{-1}),
\end{align}
as shown in Appendix \ref{app: overlap}.
This is derived from the unperturbed ground state \(\ket{E_0^{(0)}}\), which is the fully polarized state with spins aligned in the negative \(y\)-axis direction, and the prediction for \(\ket{E_1}\):
\begin{align}
    \Ket{E_1}
    \approx&
    \sum_{k=1}^L
    \frac{(-1)^{k-1}}{\sqrt{L/2}}
    \sin\left(\frac{\pi}{L}\left(k-\frac{1}{2}\right)\right)
    \sigma^x_k\Ket{E_0^{(0)}}.
\end{align}
In between, \(0<B_y<\infty\), exact diagonalization tells us \(|\Braket{E_1|E_0'}|^2 > \frac{8}{\pi^2}=81\%\) over a wide range of \(B_y\) irrespective of \(L\) (see Appendix \ref{app: overlap}).
% ~\cite{sm}. 
Remarkably, the wave function overlap, and thus the DTC-like signal, does not vanish even as $L$ increases.  These observations show that the behavior of Eq.~\eqref{eq:main} originates from Eq.~\eqref{eq:E1approx} in our model.

Notably, Eq.~\eqref{eq:E1approx} poses a duality between the UPDTC and a PDTC with polarization as follows. 
%To see this, we introduce two notions.
First, we define, for some real $\phi$, the following generalized N\'{e}el states,
\begin{align}
    \Ket{\tilde{\pm},\phi}
    =
    \frac{\Ket{E_0}\pm e^{i\phi} \Ket{E_0'}}{\sqrt{2}}
    \approx
    \frac{\Ket{E_0}\pm e^{i\phi} \Ket{E_1}}{\sqrt{2}},
\end{align}
which have staggered polarization.
In fact, \(\Ket{\tilde{\pm},\phi}\) is not invariant under \(P_{\pi}^{(y)}\), as shown by the relation \(P_{\pi}^{(y)} \Ket{\tilde{\pm},\phi} = i^L \Ket{\tilde{\mp},\phi}\). For \(B_y = 0\), \(\Ket{\tilde{\pm}, 0}\) coincide with the N\'{e}el states. 
%The autocorrelation, \(C(nT)\), can also be intuitively understood as the staggered magnetization for a certain state exhibiting staggered polarization. In fact, 
Second, \(C(nT)\) is related to the staggered magnetization expectation value as
\begin{align}
    \label{eq:Cmag}
    \mathrm{Re}\left(e^{i\phi}C(nT)\right)
    =
    \frac{\Braket{M_x^{\mathrm{st}}(nT)}_{\tilde{+},\phi}-\Braket{M_x^{\mathrm{st}}(nT)}_{\tilde{-},\phi}}{2\mathcal{N}},
\end{align}
where \(\Braket{M_x^{\mathrm{st}}(nT)}_{\tilde{\pm},\phi}\equiv \Bra{\tilde{\pm},\phi}M_x^{\mathrm{st}}(nT) \Ket{\tilde{\pm},\phi}\).
%Therefore, \(C(nT)\) can be interpreted as the dynamics of the staggered magnetization of the state created by the superposition of the ground and first excited states.
These two notions provide the dual picture as summarized in Fig.~\ref{fig:main-results}(c): The same DTC-like signal is also obtained by preparing $\Ket{\tilde{\pm},\phi}$ and measuring the staggered magnetization rather than its autocorrelation.
%The time evolution, obtained by sequentially applying \(U_1\) and \(U_2\) to \(\Ket{\tilde{\pm}, \phi}\),
In this picture, \(\Ket{\tilde{\pm}, \phi}\) oscillates back and forth with a period of \(2T\), but \(\phi\) shifts by \(-2T\Delta E_{01}\) according to the energy difference. We emphasize again that, in our original picture, the state is not polarized, but the DTC-like signal is obtained by measuring the autocorrelation instead of the staggered magnetization.

\section{Experimental proposal \label{sec: experiment}} 
% \lsec{Experimental proposal} 
Finally, we discuss a setup for experimentally observing UPDTC. Since our model is inspired by ion trap experiments \cite{Kyprianidis2021}, UPDTCs are expected to be observed primarily on such a platform. Specifically, we propose measuring \(\mathrm{Im}C(nT)\) in the paramagnetic phase. To perform the measurement, preparing the UGS \(\Ket{E_0}\) is first necessary. For very large \(B_y\), \(\Ket{E_0}\) can be approximately described as the fully polarized state pointing in the negative \(y\)-direction. Starting with this initial state, \(\Ket{E_0}\) can be obtained by gradually lowering \(B_y\) to the target value following the method of adiabatic state preparation. The following relation can be used to measure \(\mathrm{Im}C(nT)\):
\begin{align}
    \label{eq:ImC}
    &\mathrm{Im}C(nT)
    \notag
    \\
    &
    =
    \sum_{j=1}^L \frac{(-1)^{j-1}}{\mathcal{N}^2}
    \left(
    \Braket{M_{x}^{\mathrm{st}}(nT)}^{\frac{\pi}{4}}_{j}
    -
    \Braket{M_{x}^{\mathrm{st}}(nT)}^{-\frac{\pi}{4}}_{j}
    \right),
\end{align}
where $\Braket{\cdots}_{j}^{\pm\frac{\pi}{4}}$ represents the expectation value for the states $R^x_j(\pm \pi/4)\Ket{E_0}$ with \(R^x_j(\theta) = e^{-i(\theta/2)\sigma^x_j}\).
Thus, by applying \(R^x_j(\pm \pi/4)\) to the prepared \(\Ket{E_0}\), evolving the resulting state in time, and measuring the staggered magnetization, \(\mathrm{Im}C(nT)\) can be determined through Eq.~\eqref{eq:ImC}. 

\section{Conclusions \label{sec: conclusions}}
% \lsec{Conclusions}
We have unveiled a new class of prethermal discrete time crystals, termed unpolarized prethermal DTCs, that exhibit time-crystalline order without relying on polarization. By studying a periodically driven spin system, we showed that quantum fluctuations of the staggered magnetization can reveal coherent dynamics and DTC-like signatures (Eqs.~\eqref{eq:main} and \eqref{eq:main_rot}), even when the order parameter itself vanishes. This unpolarized phase is exponentially long-lived in the high-frequency regime, a hallmark of prethermalization. Our results expand the landscape of prethermal time crystals and highlight the intricate interplay between quantum effects and symmetries in stabilizing novel out-of-equilibrium phases.

Looking forward, our work opens up exciting avenues for future research, including exploring unpolarized time crystals in higher dimensions, different lattice geometries, and systems with disorder. Experimentally, our proposed scheme to observe unpolarized prethermal DTCs in trapped ion systems could be readily implemented, providing a platform for studying nonequilibrium quantum dynamics and the mechanisms underlying robust time-crystalline order.

\acknowledgments
% \lsec{Acknowledgments}
T.N.I. thanks A.~Polkovnikov and S.~Sugiura for collaborating on previous related studies.
Numerical exact diagonalization in this work has been performed with the help of the \texttt{QuSpin} package~\cite{Weinberg2017,Weinberg2019}, and quantum time evolution with \texttt{Qulacs}~\cite{Suzuki2021}.
This work was supported by JST PRESTO Grant No. JPMJPR2112 and by JSPS KAKENHI Grant Nos. JP21K13852, 25K07178, 25H00177, and 25K23364.

\appendix

\section{Transition point in the Floquet effective Hamiltonian \label{app: transition point}}
The leading-order Floquet effective Hamiltonian \(H_{\mathrm{eff}}^{\mathrm{(0)}}\) corresponds to the transverse field Ising model, which undergoes an antiferromagnetic phase transition \cite{Koffel2012}, with the presence or absence of ground-state degeneracy changing across the critical field \(B_y = B_{\mathrm{c}}\). To understand the Floquet dynamics discussed in the main text, we estimate \(B_{\mathrm{c}}\) by calculating the energy difference \(\Delta E_{01}\) between the two lowest eigenvalues using exact diagonalization for various \(L\). 

\begin{figure}
    \centering
    \includegraphics[width=\linewidth]{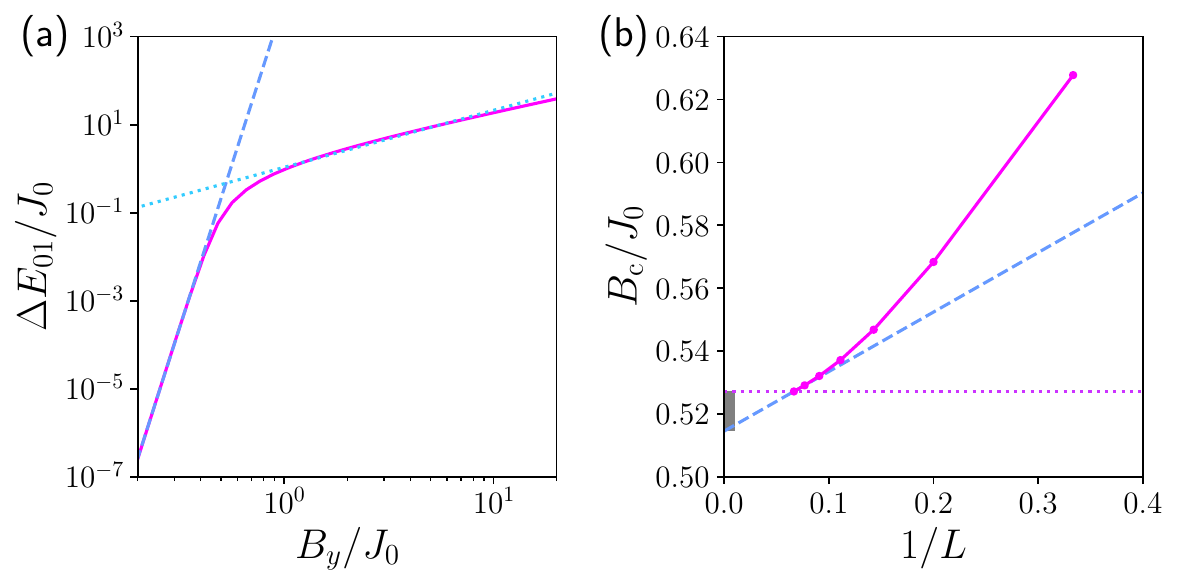}
    \caption{Results regarding the estimates of \(B_c/J_0\). (a) \(B_y\) dependence of \(\Delta E_{01}\) at \(L=15\). The dashed line (dotted line) represents the linear fitting of \(\Delta E_{01}\) in the range \(0.2 \leq B_y/J_0 \leq 0.4\) (\(0.8 \leq B_y/J_0 \leq 10.0\)) in a log-log scale. (b) \(1/L\) dependence of the estimated values of \(B_c/J_0\) for \(L\leq 15\). The magenta dashed line represents the value at \(L=15\), while the blue dashed line is a linear fit to the two leftmost points. The gray region indicates the range between the \(1/L\to 0\) values of these two lines.}
    \label{fig:trans_point}
\end{figure}

Figure \ref{fig:trans_point}(a) shows the \(B_y\)-dependence of \(\Delta E_{01}\) in the case of \(L=15\). The results indicate that \(\Delta E_{01}\) decreases rapidly across a specific region as the field is reduced. In this figure, the results are linearly fitted in regions of small and large \(B_y\), and the intersection of these linear fits is used to estimate \(B_{\mathrm{c}}\). 

Similar analyses are conducted for various values of \(L\) to investigate the thermodynamic limit. Figure \ref{fig:trans_point}(b) plots the estimates of \(B_c/J_0\) for \(L\leq 15\). The magenta dotted line represents the value at \(L=15\), while the blue dashed line is a linear fit to the two leftmost points. From this figure, the estimated value of \(B_c/J_0\) increases monotonically and exhibits convex behavior as a function of \(1/L\). Thus, it is inferred that \(B_c/J_0\) converges to the gray-shaded region in the figure as \(L\to\infty\). Based on this observation, we estimate the transition point in the thermodynamic limit as \(B_c/J_0=0.52\pm 0.01\), which is consistent with the transition point suggested in Ref.~\cite{Koffel2012} for \(L \to \infty\), indicating a phase transition near \(B_{\mathrm{c}}/J_0 \sim 0.5\).

\section{Calculation method for time evolution \label{app: method time evolution}}

We describe our numerical calculation method for the autocorrelation function \(C(nT)\) for a state \(\Ket{\psi}\).

The autocorrelation function can be expressed as follows:
\begin{align}
    C(nT)=\frac{1}{\mathcal{N}}\Braket{\psi(n)|M_x^{\mathrm{st}}|\psi'(n)}.
\end{align}
Here, the states \(\Ket{\psi(n)}\) and \(\Ket{\psi'(n)}\) are defined as follows:
\begin{align}
    \Ket{\psi(n)}=&(U_2 U_1)^{n}\Ket{\psi},
    \\
    \Ket{\psi'(n)}=&(U_2 U_1)^{n}\frac{M_x^{\mathrm{st}}\Ket{\psi}}{\|M_x^{\mathrm{st}}\Ket{\psi}\|}.
\end{align}
By calculating the time evolution of \(\Ket{\psi(n)}\) and \(\Ket{\psi'(n)}\), we obtain \(C(nT)\).

The states' time evolution is calculated using a quantum circuit simulator after applying the Trotter decomposition. First, \(U_1\) is divided into \(M\) factors:
\begin{align}
    U_1=&u_1(\delta)^M,
    \\
    u_1(\delta)
    =&
    e^{-i\delta\left(\sum_{i<j}^{N}J_{ij}\sigma^x_i \sigma^x_j + B_y\sum_{i=1}^N\sigma^y_i+B_z\sum_{i=1}^N\sigma^z_i\right)},
\end{align}
where \(\delta = T / M\). We apply the second-order Trotterization formula to \(u_1(\delta)\), achieving a decomposition with an accuracy of \(O(\delta^2)\).
\begin{widetext}
\begin{align}
    \label{eq:u1_gate}
    u_1(\delta)
    =&
    \left(\prod_{i=1}^L e^{-i\frac{\delta}{2}B_z\sigma^z_i}\right)
    \left(\prod_{i=1}^L e^{-i\frac{\delta}{2}B_y\sigma^y_i}\right)
    % \notag
    % \\
    % &\times
    \left(\prod_{i<j}^L e^{-i\delta J_{ij}\sigma^x_i\sigma^y_i}\right)
    % \notag
    % \\
    % &\times
    \left(\prod_{i=1}^L e^{-i\frac{\delta}{2}B_y\sigma^y_i}\right)
    \left(\prod_{i=1}^L e^{-i\frac{\delta}{2}B_z\sigma^z_i}\right)
    +
    O\left(\delta^3\right).
\end{align}
In terms of quantum gates, this can be expressed as follows:
\begin{align}
    u_1(\delta)
    =&
    \left(\prod_{i=1}^L \mathrm{RZ}_i(B_z\delta)\right)
    \left(\prod_{i=1}^L \mathrm{RY}_i(B_y\delta)\right)
    \left(\prod_{i<j}^L 
    \mathrm{RY}_i\left(\frac{\pi}{2}\right)
    \cdot
    \mathrm{CZ}_{ij}
    \cdot
    \mathrm{RX}_{j}(2\delta J_{ij})
    \cdot
    \mathrm{CZ}_{ij}
    \cdot
    \mathrm{RY}_i\left(-\frac{\pi}{2}\right)
    \right)
    \notag
    \\
    &\times
    \left(\prod_{i=1}^L \mathrm{RY}_i(B_y\delta)\right)
    \left(\prod_{i=1}^L \mathrm{RZ}_i(B_z\delta)\right)
    +
    O\left(\delta^3\right).
\end{align}
\end{widetext}
Here, \(\mathrm{CZ}_{ij}\) represents the controlled-$Z$ gate, and the rotation gates are defined as follows.
\begin{align}
    \mathrm{RX}_i(\theta)=e^{-i\frac{\theta}{2}\sigma_x^i},\,
    \mathrm{RY}_i(\theta)=e^{-i\frac{\theta}{2}\sigma_y^i},\,
    \mathrm{RZ}_i(\theta)=e^{-i\frac{\theta}{2}\sigma_z^i}.
\end{align}
To derive Eq.~\eqref{eq:u1_gate}, the following relation has been used:
\begin{align}
    &e^{-i\delta J_{ij} \sigma^x_i \sigma^x_j}
    \notag
    \\
    &=
    \mathrm{RY}_i\left(\frac{\pi}{2}\right)
    \cdot
    \mathrm{CZ}_{ij}
    \cdot
    \mathrm{RX}_{j}(2\delta J_{ij})
    \cdot
    \mathrm{CZ}_{ij}
    \cdot
    \mathrm{RY}_i\left(-\frac{\pi}{2}\right).
\end{align}
Additionally, \(U_2\) is expressed in terms of quantum gates as
\begin{align}
    U_2=\prod_{i=1}^{L}\mathrm{RY}_{i}(\pi).
\end{align}
By repeatedly applying the above quantum gates, we calculate \(\Ket{\psi(n)}\) and \(\Ket{\psi'(n)}\). In this work, we consider \(\Ket{\psi}\) as the UGS \(\Ket{E_0}\) and the N\'eel state \(\Ket{\mathrm{NS}+}\). We prepare
\(\Ket{E_0}\) using exact diagonalization.

\begin{figure}
    \centering
    \includegraphics[width=\linewidth]{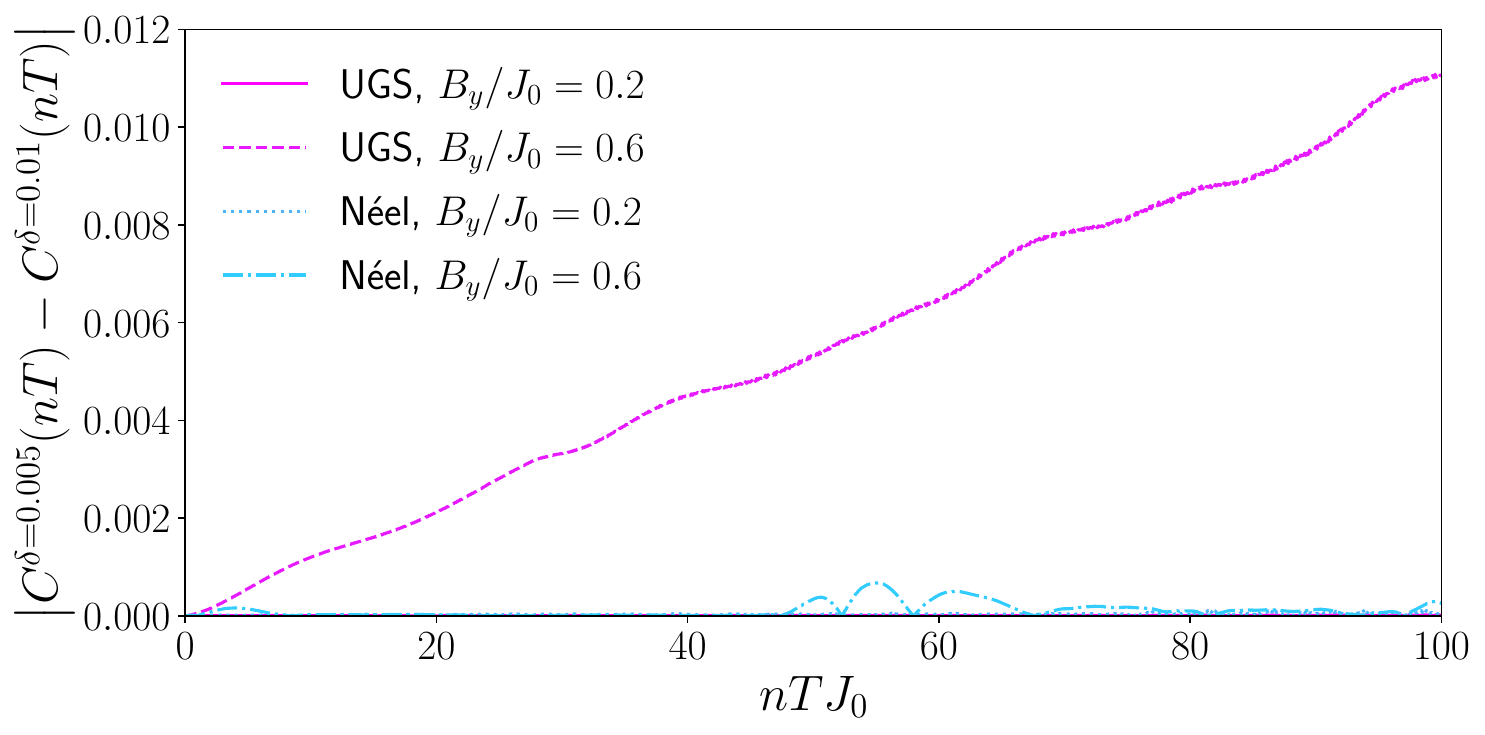}
    \caption{Deviation of \(C(nT)\) for \(\delta = 0.005\) and \(\delta = 0.01\) for the UGS and N\'eel states under \(B_y/J_0 = 0.2\) and \(B_y/J_0 = 0.6\) at \(TJ_0 = 0.1\).}
    \label{fig:Merr}
\end{figure}
We fix \(\delta=0.01\) and set \(M\) as \(M=\lceil T/\delta \rceil\) in the calculation. We discuss the convergence of the results with respect to the Trotter error. Figure \ref{fig:Merr} plots the deviation \(|C^{\delta=0.005}(nT) - C^{\delta=0.01}(nT)|\) for both the UGS and N\'eel states under the parameters used in Fig.~1(b). Here, \(C^{\delta=0.005}(nT)\) and \(C^{\delta=0.01}(nT)\) denote the results for \(\delta=0.005\) and \(\delta=0.01\), respectively. This figure shows that even for the UGS result at \(B_y/J_0=0.6\), where the deviation is the largest, the discrepancy relative to the typical signal size \(C(0)=1\) remains within 1.2\%.

\section{Results for short-range interactions \label{app: short range}}
\begin{figure*}
    \centering
    \includegraphics[width=\linewidth]{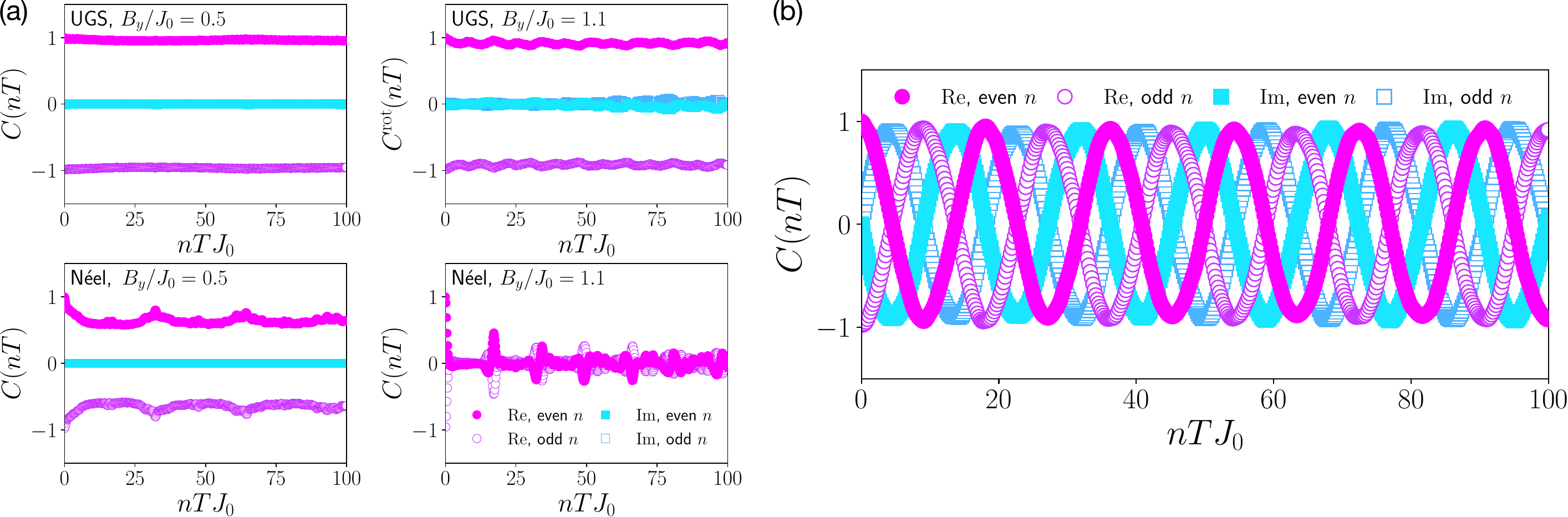}
    \caption{Results for $L = 15$ and \(TJ_0=0.1\) with nearest-neighbor interactions. (a) \(C(nT)\) and \(C^{\mathrm{rot}}(nT)\) for the UGS and N\'eel state  at \(B_y/J_0 = 0.5\) (antiferromagnetic phase) and \(B_y/J_0 = 1.1\) (paramagnetic phase). (b) \(C(nT)\) of the UGS at \(B_y/J_0 = 1.1\).}
    \label{fig:Ising}
\end{figure*}
We discuss how the presence of the UPDTC depends on the interaction range, focusing in particular on the case of nearest-neighbor interactions $J_{ij} = J_0 \delta_{|i-j|,1}$.

Figure~\ref{fig:Ising} shows the results for $L = 15$ and \(TJ_0=0.1\) with nearest-neighbor interactions. In this case, the Floquet effective Hamiltonian corresponds to the transverse-field Ising model, which exhibits a phase transition at $B_y/J_0 = B_c/J_0 = 1$. Figure~\ref{fig:Ising}(a) presents the results of $C(nT)$ and $C^{\mathrm{rot}}(nT)$ for the UGS and N\'eel states at $B_y/J_0 = 0.5$ (antiferromagnetic phase) and $B_y/J_0 = 1.1$ (paramagnetic phase). Here, $C^{\mathrm{rot}}(nT)$ is derived from $C(nT)$ shown in Fig.~\ref{fig:Ising}(b). These results are qualitatively similar to those for long-range interactions (Fig.~1(b) and Fig.~2 in the main text). Namely, in contrast to the N\'eel state, the UGS exists regardless of $B_y/J_0$, and for $B_y > B_c$, $C(nT)$ exhibits a sinusoidal signal. Therefore, in our setup, the qualitative properties of the UPDTC are independent of the interaction range.

It is interesting to discuss how these results hold in the thermodynamic limit, where the effects of the interaction range become more pronounced. For short-range interactions, finite-temperature phase transitions are forbidden by the Landau-Peierls argument \cite{Landau1937}, so SSB PDTC is believed to be unable to survive in the thermodynamic limit \cite{Machado2020,Kyprianidis2021}. 
However, since UPDTC is not based on SSB, the Landau-Peierls argument does not rule out its existence. Clarifying whether UPDTC survives in the thermodynamic limit remains a task for future research.

\section{Quantification of the signal lifetime extension \label{app: lifetime}}
\begin{figure}
    \centering
    \includegraphics[width=\linewidth]{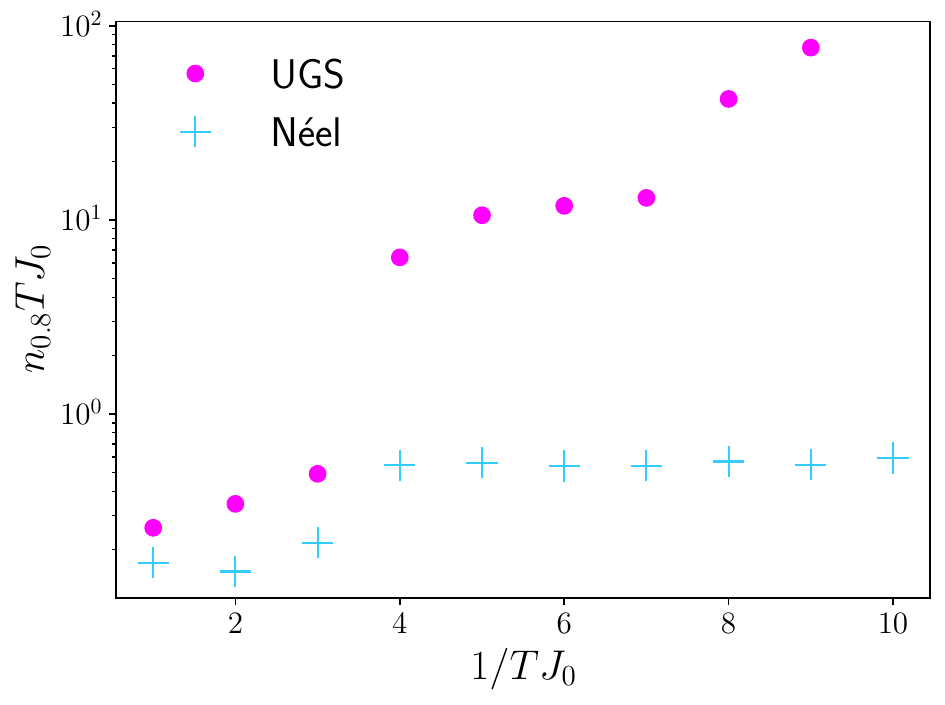}
    \caption{DTC-like signal's lifetime, \(n_{0.8}TJ_0\), as a function of drive frequency \(1/TJ_0\) under \(B_y/J_0=0.6\) for the UGS (circle) and N\'eel state (plus).}
    \label{fig:Lifetime}
\end{figure}
We quantify the lifetimes of the signals shown in Fig.~\ref{fig:absC}
% Fig.~3 
of the main text by extracting the value of $n$ at which $|C(nT)|$ reaches $0.8$, denoted as $n_{0.8}$. Figure~\ref{fig:Lifetime} displays the dependence of the time $n_{0.8}TJ_0$ on the drive frequency for the UGS and N\'eel state at $B_y/J_0 = 0.6$. This indicates that the signal lifetime for the UGS increases exponentially with the drive frequency, in contrast to the N\'eel state.

\section{Robustness of the signal against noise \label{app: robustness}}
\begin{figure}
    \centering
    \includegraphics[width=\linewidth]{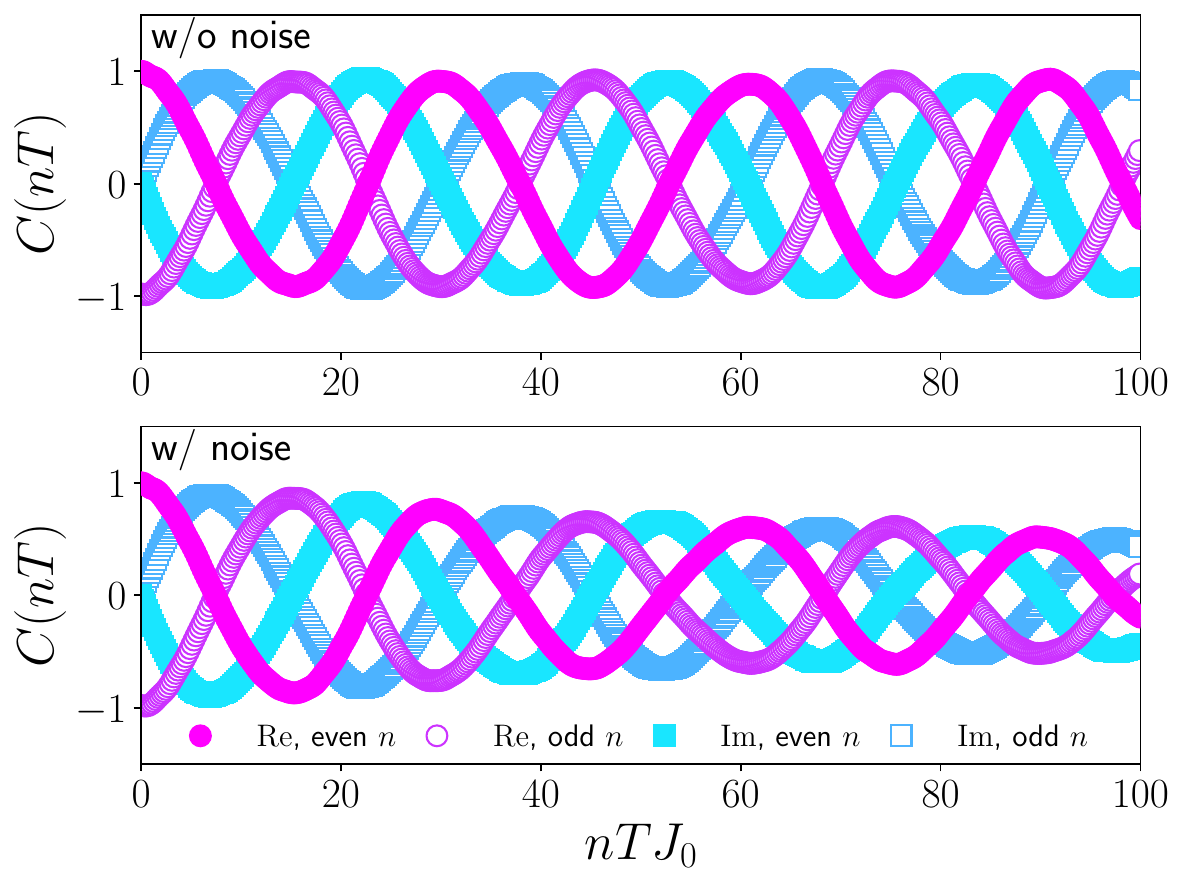}
    \caption{Signal at $L=15$ with and without noise in the magnetic field. The magnetic field is set to $B_y/J_0 = 0.6$ and $B_z/J_0 = 1.0$. In the noisy case, the magnetic field is replaced by $(1+\epsilon_y)B_y$ and $(1+\epsilon_z)B_z$, where $\epsilon_{y,z}$ is independently and randomly chosen from $[-0.5, 0.5]$ for each quantum gate and each Trotter step}.
    \label{fig:Noise}
\end{figure}
We discuss the robustness of the signal against noise, focusing on how the signal changes in the presence of noise in the magnetic field. Figure~\ref{fig:Noise} shows the results of $C(nT)$ with and without noise at $L = 15$, $B_y/J_0 = 0.6$, and $B_z/J_0 = 1.0$. In the noisy case, the magnetic field is replaced by $(1+\epsilon_y)B_y$ and $(1+\epsilon_z)B_z$, where $\epsilon_{y,z}$ is independently and randomly chosen from $[-0.5, 0.5]$ for each quantum gate and each Trotter step. This figure shows that, in the presence of noise, the amplitude of $C(nT)$ decays over time, while its overall structure remains unchanged.

This behavior is similar to that of the SSB PDTC. In fact, simulations of the SSB PDTC using the Krylov subspace method, presented in the supplemental material of Ref.~\cite{Kyprianidis2021}, also show a decay in the signal amplitude over time due to noise. We emphasize that this does not imply the signal becomes unobservable in experiments. Indeed, in Ref.~\cite{Kyprianidis2021}, a finite-lifetime SSB PDTC signal is observed, even though the lifetime extension with increasing drive frequency is suppressed by noise.

\section{Deviation of the peak position in the Fourier component of \(C(nT)\) \label{app: peak}}
As discussed in the main text, the frequency of the phase change \(\Omega\) exhibited by the UPDTC is close to \(\Delta E_{01}\) obtained from \(H_{\mathrm{eff}}^{(0)}\), but there is a slight deviation. Here, we discuss the origin of the deviation.

Table \ref{tab:omega_deviation} shows the results for \(\Omega\) and \(\Delta E_{01}\) for several \(B_y\) values at \(TJ_0 = 0.1\) and \(0.05\). These results show that as \(T\) decreases, the difference between \(\Omega\) and \(\Delta E_{01}\) decreases. Since the system can be described more accurately by \(H_{\mathrm{eff}}^{(0)}\) as \(T\) decreases, our results suggest that the difference between \(\Omega\) and \(\Delta E_{01}\) arises from the use of an approximate effective Hamiltonian \(H_{\mathrm{eff}}^{(0)}\).
\begin{table}[h]
    \caption{\(\Delta E_{01}/J_0\) and \(\Omega/J_0\) for \(TJ_0 = 0.1, 0.05\) in the cases of \(B_y/J_0 = 0.6, 5.0, 10.0\). The results of \(\Omega/J_0\) are calculated from \(C(nT)\) within the range \(0 \leq nTJ_0 \leq 100\).\label{tab:omega_deviation}}
    \centering
    \begin{tabular}{c|cccc}
        \hline\hline
        \(B_y/J_0\)  & 0.6  & 5.0  & 10.0 \\
        \hline
        \(\Omega/J_0\) (\(TJ_0=0.10\)) & 0.186 & 8.69 & 18.5  \\
        \(\Omega/J_0\) (\(TJ_0=0.05\)) & 0.204 & 8.75 & 18.7 \\
        \hline
        \(\Delta E_{01}/J_0\) & 0.210 & 8.76 & 18.7
        \\
        \hline\hline
    \end{tabular}
\end{table}

\begin{widetext}
\section{Derivation of Eq.~\eqref{eq:Cspectral} \label{app: derivation}}
Our derivation of Eq.~\eqref{eq:Cspectral} utilizes the decomposition of \((U_2 U_1)^n\) via the Baker–Campbell–Hausdorff formula. Using the fact that \(U_1\) can be expressed in terms of \(H_{\mathrm{eff}}^{(0)}\) and \(M_z = \sum_{i=1}^L \sigma^z_i\) as
\begin{align}
    U_1=e^{-iT(H_{\mathrm{eff}}^{(0)}+B_z M_z)},
\end{align}
\(U_2 U_1\) can be rewritten as
\begin{align}
    U_2 U_1
    =
    P_\pi^{(y)}e^{-iT(H_{\mathrm{eff}}^{(0)}+B_z M_z)}
    =
    e^{-iT(H_{\mathrm{eff}}^{(0)}-B_z M_z)} P_\pi^{(y)}
    =
    e^{iTB_z M_z} e^{-iTH_{\mathrm{eff}}^{(0)}}P_\pi^{(y)} + O(T^2).
\end{align}
Additionally, \((U_2 U_1)^2\) can be rewritten as
\begin{align}
    (U_2 U_1)^2
    =&
    P_\pi^{(y)}e^{-iT(H_{\mathrm{eff}}^{(0)}+B_z M_z)}
    P_\pi^{(y)}e^{-iT(H_{\mathrm{eff}}^{(0)}+B_z M_z)}
    =
    e^{-iT(H_{\mathrm{eff}}^{(0)}-B_z M_z)}
    e^{-iT(H_{\mathrm{eff}}^{(0)}+B_z M_z)}
    P_\pi^{(y)2}
    \notag
    \\
    =&
    e^{-i2TH_{\mathrm{eff}}^{(0)}}
    P_\pi^{(y)2}
    +
    O(T^2).
\end{align}
Repeatedly applying the above two results, we obtain 
\begin{align}
    (U_2 U_1)^n=e^{i\frac{1-(-1)^{n}}{2} TB_z M_z}e^{-inTH_{\mathrm{eff}}^{(0)}}P_\pi^{(y)n}  + O(T^2).
\end{align}

Using this approximation, \(M_x^{\mathrm{st}}(nT)\) can be evaluated as follows:
\begin{align}
    M_x^{\mathrm{st}}(nT)
    =&
    (U_2 U_1)^{\dagger n} M_x^{\mathrm{st}} (U_2 U_1)^{n}
    =
    P_\pi^{(y)\dagger n}
    e^{inTH_{\mathrm{eff}}^{(0)}}
    e^{-i\frac{1-(-1)^{n}}{2} TB_z M_z}
    M_x^{\mathrm{st}}
    e^{i\frac{1-(-1)^{n}}{2} TB_z M_z}e^{-inTH_{\mathrm{eff}}^{(0)}}P_\pi^{(y)n}  + O(T^2)
    \notag
    \\
    =&
    P_\pi^{(y)\dagger n}
    e^{inTH_{\mathrm{eff}}^{(0)}}
    \sum_{i=1}^L
    (-1)^{i-1}
    e^{-i\frac{1-(-1)^{n}}{2} TB_z \sigma^z_i}
    \sigma^x_i
    e^{i\frac{1-(-1)^{n}}{2} TB_z \sigma^z_i}
    e^{-inTH_{\mathrm{eff}}^{(0)}}P_\pi^{(y)n}
    +
    O(T^2)
    \notag
    \\
    =&
    \cos \left((1-(-1)^{n})TB_z\right)
    P_\pi^{(y)\dagger n}
    e^{inTH_{\mathrm{eff}}^{(0)}}
    M_x^{\mathrm{st}}
    e^{-inTH_{\mathrm{eff}}^{(0)}}P_\pi^{(y)n}
    \notag
    \\
    &+
    \sin \left((1-(-1)^{n})TB_z\right)
    P_\pi^{(y)\dagger n}
    e^{inTH_{\mathrm{eff}}^{(0)}}
    M_y^{\mathrm{st}}
    e^{-inTH_{\mathrm{eff}}^{(0)}}
    P_\pi^{(y)n}
    +
    O(T^2)
    \notag
    \\
    =&
    (-1)^n
    e^{inTH_{\mathrm{eff}}^{(0)}}
    M_x^{\mathrm{st}}
    e^{-inTH_{\mathrm{eff}}^{(0)}}
    +
    \sin \left((1-(-1)^{n})TB_z\right)
    e^{inTH_{\mathrm{eff}}^{(0)}}
    M_y^{\mathrm{st}}
    e^{-inTH_{\mathrm{eff}}^{(0)}}
    +
    O(T^2).
\end{align}
Here, \(M_y^{\mathrm{st}} = \sum_{i=1}^L (-1)^{i-1}\sigma^y_i\) is introduced, and the relations \(\{P_\pi^{(y)}, M_x^{\mathrm{st}}\} = 0\) and \([P_\pi^{(y)}, M_y^{\mathrm{st}}] = 0\) are used.

Using the results above, \(C(nT)\) can be evaluated as
\begin{align}
    \label{eq:CnTeval}
    &C(nT)
    =
    \frac{1}{\mathcal{N}^2}
    \Braket{E_0|M_x^{\mathrm{st}}(nT)M_x^{\mathrm{st}}|E_0}
    \notag
    \\
    =&
    \frac{(-1)^n}{\mathcal{N}^2}
    \Braket{E_0|
    e^{inTH_{\mathrm{eff}}^{(0)}}M_x^{\mathrm{st}}
    e^{-inTH_{\mathrm{eff}}^{(0)}}M_x^{\mathrm{st}}
    |E_0}
    +
    \frac{\sin \left((1-(-1)^{n})TB_z\right)}{\mathcal{N}^2}
    \Braket{E_0|
    e^{inTH_{\mathrm{eff}}^{(0)}}M_y^{\mathrm{st}}
    e^{-inTH_{\mathrm{eff}}^{(0)}}M_x^{\mathrm{st}}
    |E_0}
    +
    O(T^2)
\end{align}
Here we remark \(P_\pi^{(y)}\Ket{E_0} = i^L \Ket{E_0}\), meaning $\Ket{E_0}$ is an eigenstate of \(P_\pi^{(y)}\). Thus, it gives zero expectation value for any operator \(A\) such that \(\{A, P_\pi^{(y)}\} = 0\). In fact, we have
\begin{align}
    \label{eq:Aave}
    \Braket{E_0|A|E_0}
    =
    \Braket{E_0|P_\pi^{(y)\dagger}P_\pi^{(y)}A|E_0}
    =
    -\Braket{E_0|P_\pi^{(y)\dagger}AP_\pi^{(y)}|E_0}
    =
    -
    \Braket{E_0|A|E_0}
    =
    0
\end{align}
This implies that the second term on the rightmost side of Eq.~\eqref{eq:CnTeval} vanishes, leading to
\begin{align}
    C(nT)
    =&
    \frac{(-1)^n}{\mathcal{N}^2}
    \Braket{E_0|
    e^{inTH_{\mathrm{eff}}^{(0)}}M_x^{\mathrm{st}}
    e^{-inTH_{\mathrm{eff}}^{(0)}}M_x^{\mathrm{st}}
    |E_0}
    +
    O(T^2).
\end{align}
By inserting the completeness relation \(\sum_{m} \Ket{E_m} \Bra{E_m} = 1\) into this equation, and using \(\Braket{E_0 | M_x^{\mathrm{st}} | E_0} = 0\), which follows from Eq.~\eqref{eq:Aave}, we obtain
\begin{align}
    C(nT)=\frac{(-1)^n}{\mathcal{N}^2}\sum_{m\geq 1}e^{-inT(E_{m}-E_0)} |\Braket{E_m|M_x^{\mathrm{st}}|E_0}|^2+O(T^2).
\end{align}
Using \(\mathcal{N}^2=\Braket{M_x^{\mathrm{st}2}}=\|M_x^{\mathrm{st}} \Ket{E_0}\|^2\), we obtain Eq.~\eqref{eq:Cspectral}.
\end{widetext}

\section{Overlap between \texorpdfstring{$\Ket{E_0'}$}{E0'} and \texorpdfstring{$\Ket{E_1}$}{E1} \label{app: overlap}}
\begin{figure*}
    \centering
    \includegraphics[width=\linewidth]{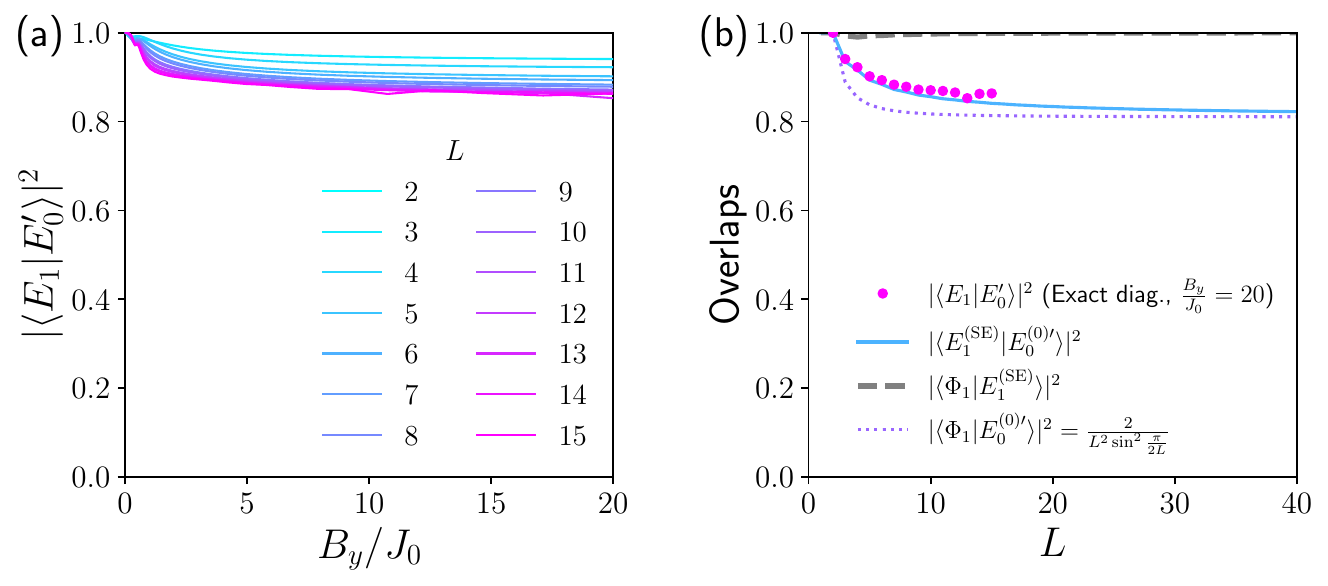}
    \caption{Results about overlaps between states. (a) \(B_y/J_0\) dependence of \(|\Braket{E_1|E_0'}|^2\) for various \(L\) obtained via the exact diagonalization. (b) \(L\) dependence of \(|\Braket{E_1|E_0'}|^2\) obtained by the exact diagonalization at \(B_y/J_0=20\) and overlaps among the first excited state from the secular equation \(\ket{E_1^{(\mathrm{SE})}}\), \(\ket{\Phi_1}=\sum_{k=1}^L\phi_1(k)\ket{E_{1,k}^{(0)}}\), and \(\ket{E_0^{(0) \prime}} = M_x^{\mathrm{st}} \ket{E_0^{(0)}}/\| M_x^{\mathrm{st}} \ket{E_0^{(0)}}\|\).}
    \label{fig:overlap}
\end{figure*}
We discuss the overlap between \(\Ket{E_0'}\) and \(\Ket{E_1}\) under the leading-order Floquet effective Hamiltonian \(H_{\mathrm{eff}}^{(0)}\). As stated, \(|\Braket{E_1|E_0'}|^2=1\) holds exactly for \(B_y = 0\). Here, we first present exact diagonalization results for \(B_y > 0\), and then interpret the observed behavior for \(B_y/J_0 \gg 1\) using perturbative analysis.

Figure \ref{fig:overlap}(a) shows \(|\Braket{E_1|E_0'}|^2\) calculated via exact diagonalization for various \(L\) and \(B_y\). This figure shows that, for each \(L\), \(|\Braket{E_1|E_0'}|^2\) decreases to converge to a certain value as \(B_y\) increases. In addition, \(|\Braket{E_1|E_0'}|^2\) also decreases to converge to a nonzero value when $L$ increases.
Consequently, for all parameter ranges examined, \(|\Braket{E_1|E_0'}|^2 > 80\%\) is satisfied.

To understand the behaviors of \(|\Braket{E_1|E_0'}|^2\), we consider the case \(B_y/J_0 \gg 1\) and perform a perturbative analysis with the following choices for the unperturbed Hamiltonian \(H_{0}\) and the perturbation \(H_{1}\):
\begin{align}
    H_{\mathrm{eff}}^{\mathrm{(0)}}=&H_0+H_1,
    \\
    H_0=&\sum_{i=1}^L B_y \sigma^y_i,
    \\
    \label{eq:H1}
    H_1=&\sum_{i<j}J_{ij}\sigma^x_i \sigma^x_j.
\end{align}
As in the main text, for \(i \neq j\), we define \(J_{ij} = J_0 / |i - j|\) (\(J_0>0)\)), and additionally, we set \(J_{ii} = 0\). We analyze the ground state and the first excited state to calculate \(|\Braket{E_1|E_0'}|^2\). The ground state of \(H_{0}\) is given by 
\begin{align}
    \Ket{E_0^{(0)}}=\Ket{\leftarrow}_1\otimes \cdots\otimes \Ket{\leftarrow}_L,
\end{align}
where \(\ket{\leftarrow}_i\) represents the spin at site \(i\) pointing in the negative \(y\)-direction. The first excited state of \(H_{0}\) is obtained by flipping the spin of one site in \(\ket{E_0^{(0)}}\). Therefore, it forms a degenerate set of \(L\) states, as follows:
\begin{align}
    \label{eq:E10}
    \Ket{E_{1,k}^{(0)}}= \sigma^{x}_k \Ket{E_0^{(0)}}\quad (k=1,\ldots,L).
\end{align}
This degeneracy is lifted by the perturbation \(H_{1}\), and the first excited state can be expressed as
\begin{align}
    \Ket{E_1}
    =
    \sum_{k=1}^L a_k \Ket{E_{1,k}^{(0)}}+O(J_0/B_y),
\end{align}
where the coefficients \(a_k\) are determined as the normalized eigenvector corresponding to the smallest eigenvalue \(\mathcal{E}\) of the secular equation
\begin{align}
    \sum_{l=1}^L
    \Braket{E_{1,k}^{(0)}|H_1|E_{1,l}^{(0)}}
    a_l
    =
    \mathcal{E}a_k.
\end{align}
By use of Eqs.~\eqref{eq:H1} and \eqref{eq:E10}, this is evaluated as follows:
\begin{align}
    \label{eq:secular}
    \sum_{l=1}^L J_{kl}a_l = \mathcal{E} a_k.
\end{align}
Hereafter, we denote the first excited state obtained from this secular equation as \(\ket{E_1^{(\mathrm{SE})}}=\sum_{k=1}^L a_k \ket{E_{1,k}^{(0)}}\), where \(a_k\) is taken as a real number because \(J_{kl}\) is a real symmetric matrix, and its eigenvectors can be chosen to be real-valued.

We calculate \(\ket{E_1^{(\mathrm{SE})}}\) via numerical diagonalization of Eq.~\eqref{eq:secular}. Figure \ref{fig:overlap}(b) shows the results of the overlap \(|\braket{E_1^{\mathrm{(SE)}}|E_0^{(0) \prime}}|^2\), where 
\begin{align}
    \Ket{E_0^{(0) \prime}} = \frac{M_x^{\mathrm{st}} \Ket{E_0^{(0)}}}{\left\| M_x^{\mathrm{st}} \Ket{E_0^{(0)}}\right\|},
\end{align}
calculated for various \(L\). This figure shows that \(|\braket{E_1^{\mathrm{(SE)}}|E_0^{(0) \prime}}|^2\) aligns well with \(|\braket{E_1|E_0'}|^2\) obtained via exact diagonalization for \(B_y/J_0 = 20\), as indicated by the magenta points. Therefore, the results at large \(B_y\) in Fig.~\ref{fig:overlap}(a) can be understood in the perturbative picture. Additionally, an important finding is that the overlap remains finite even when \(L\) is large.

The numerical diagonalization of Eq.~\eqref{eq:secular} also suggests that
\begin{align}
    \label{eq:akapprox}
    a_k
    \approx
    \phi_1(k)
    \equiv
    \sqrt{\frac{2}{L}}
    (-1)^{k-1}\sin\left(\frac{\pi}{L}\left(k-\frac{1}{2}\right)\right).
\end{align}
In fact, the overlap \(|\langle \Phi_{1}|E_1^{\mathrm{(SE)}} \rangle|^2\), where \(\ket{\Phi_1}=\sum_{k=1}^L\phi_1(k)\ket{E_{1,k}^{(0)}}\), is almost \(100\%\) regardless of \(L\), as indicated by the gray dashed line in Fig.~\ref{fig:overlap}(b). This functional form can be understood from the perspective of energy gain \(\mathcal{E}=\sum_{k=1}^L\sum_{l=1}^L J_{kl}a_k a_l\) and symmetry. Since the adjacent matrix elements satisfy \(J_{i,i+1} > 0\), a staggered configuration is energetically favored, which explains the \((-1)^{k-1}\) factor. Furthermore, the absence of neighboring sites near the edges of the spin chain enhances energy gain for configurations biased toward the chain's center compared to a uniform configuration, which is considered the origin of the sinusoidal factor. Due to the symmetry \(J_{kl}=J_{L+1-k,L+1-l}\), \(a_k\) must be either symmetric or antisymmetric under the inversion about the chain's center, \(k \to L+1-k\). This symmetry accounts for the \(-1/2\) shift in \(k\) in the sinusoidal function. An analytic derivation of Eq.~\eqref{eq:akapprox} will be presented later in this section.

From Eq.~\eqref{eq:akapprox}, it is expected that the overlap \(|\braket{E_1^{\mathrm{(SE)}}|E_0^{(0) \prime}}|^2\) can be approximately expressed by the following quantity:
\begin{align}
    \left|\Braket{\Phi_{1}|E_0^{(0)\prime}}\right|^2=\frac{2}{L^2 \sin^2 \frac{\pi}{2L}}.
\end{align}
As indicated by the purple dotted line in Fig.~\ref{fig:overlap}(b). this roughly reproduces the behavior of \(|\braket{E_1^{\mathrm{(SE)}}|E_0^{(0) \prime}}|^2\), and as \(L\) increases, the difference between the two becomes smaller. Particularly important is the behavior that \(|\braket{\Phi_{1}|E_0^{(0)\prime}}|^2\) converges to a finite value as \(L \to \infty\):
\begin{align}
    \lim_{L\to\infty}\left|\Braket{\Phi_{1}|E_0^{(0)\prime}}\right|^2=\frac{8}{\pi^2}=81\%,
\end{align}
to which \(|\braket{E_1^{\mathrm{(SE)}}|E_0^{(0) \prime}}|^2\) also shows convergence. In summary, the results from exact diagonalization and perturbation theory suggest the bound  
\begin{align}
    \left|\Braket{E_{1}|E_0^{\prime}}\right|^2 > \frac{8}{\pi^2},
\end{align}
and convergence to this lower bound as \(B_y/J_0 \to \infty\) and \(L \to \infty\).

Finally, we present an analytic derivation of Eq.~\eqref{eq:akapprox}. Motivated by the form of Eq.~\eqref{eq:akapprox}, we introduce an orthonormal basis set \(\lbrace \phi_{p}(k) \rbrace_{p=1}^L\) defined by
\begin{align}
    \label{eq:phi_k}
    \phi_p(k)
    =&
    \sqrt{\frac{2-\delta_{p,L}}{L}}
    (-1)^{k-1}\sin\left(\frac{\pi p}{L}\left(k-\frac{1}{2}\right)\right)
\end{align}
and satisfying
\begin{align}
    \sum_{k=1}^L \phi_p(k)\phi_q(k)=\delta_{p,q},\quad
    \sum_{p=1}^L \phi_p(k)\phi_p(l)=\delta_{k,l}.
\end{align}
With the expansion \(a_k=\sum_{p=1}^L \tilde{a}_p \phi_p(k)\), Eq.~\eqref{eq:secular} is rewritten as
\begin{align}
    &\sum_{q=1}^L M_{pq}\tilde{a}_q
    =
    \mathcal{E} \tilde{a}_p,
\end{align}
where
\begin{align}
    M_{pq}=\sum_{k=1}^L \sum_{l=1}^L J_{kl} \phi_{p}(k)\phi_q(l).
\end{align}
Changing the indices as \(r=k-l\) and \(n=(k+l-|r|)/2\), which take the values \(r=-L+1, -L+2,\ldots, L-1\) and \(n=1,2,\ldots,L-|r|\), introducing the notation \(J_{k-l}=J_{kl}\), and using Eq.~\eqref{eq:phi_k}, we obtain
\begin{widetext}
\begin{align}
    M_{pq}
    =&
    \sum_{r=-L+1}^{L-1} J_{r}
    \sum_{n=1}^{L-|r|}
    \phi_{p}\left(\frac{2n+|r|+r}{2}\right)\phi_{q}\left(\frac{2n+|r|-r}{2}\right)
    =
    \sum_{r=1}^{L-1} J_{r}
    \sum_{n=1}^{L-r}
    \left(
    \phi_{p}(n+r)\phi_{q}(n)
    +
    \phi_{p}(n)\phi_{q}(n+r)
    \right)
    \notag
    \\
    =&
    \frac{\sqrt{(2-\delta_{p,L})(2-\delta_{q,L})}}{L}
    \sum_{r=1}^{L-1} 
    (-1)^{r}J_{r}
    \left[
    \cos\left(\frac{\pi(p+q)r}{2L}\right)
    \sum_{n=1}^{L-r}
    \cos\left(\frac{\pi(p-q)}{L}\left(n+\frac{r-1}{2}\right)\right)
    \right.
    \notag
    \\
    &
    \qquad\qquad\qquad\qquad\qquad\qquad\qquad\quad
    -
    \left.
    \cos\left(\frac{\pi(p-q)r}{2L}\right)
    \sum_{n=1}^{L-r}
    \cos\left(\frac{\pi(p+q)}{L}\left(n+\frac{r-1}{2}\right)\right)
    \right].
\end{align}
The summations with respect to \(n\) are evaluated as
\begin{align}
    &
    \sum_{n=1}^{L-r}
    \cos\left(\frac{\pi(p-q)}{L}\left(n+\frac{r-1}{2}\right)\right)
    =
    \mathrm{Re}
    \sum_{n=1}^{L-r} e^{i\frac{\pi(p-q)}{L}\left(n+\frac{r-1}{2}\right)}
    \notag
    \\
    =&
    (L-r)\delta_{p,q}
    +
    (1-\delta_{p,q})
    \mathrm{Re}\left(
    \frac{e^{i\frac{\pi(p-q)r}{2L}}-(-1)^{p-q}e^{-i\frac{\pi(p-q)r}{2L}}}
    {e^{-i\frac{\pi(p-q)}{2L}}-e^{i\frac{\pi(p-q)}{2L}}}
    \right)
    =
    (L-r)\delta_{p,q}
    -
    (1-\delta_{p,q})\delta_{p,q}^{(\mathrm{mod}\, 2)}
    \frac{\sin\frac{\pi(p-q)r}{2L}}{\sin\frac{\pi(p-q)}{2L}},
    \\
    &
    \sum_{n=1}^{L-r}
    \cos\left(\frac{\pi(p+q)}{L}\left(n+\frac{r-1}{2}\right)\right)
    =
    \delta_{p,q}\delta_{p,L}
    (-1)^{r-1}(L-r)
    -
    (1-\delta_{p,q}\delta_{p,L})
    \delta_{p,q}^{(\mathrm{mod}\, 2)}
    \frac{\sin\frac{\pi(p+q)r}{2L}}{\sin\frac{\pi(p+q)}{2L}}
    \notag
    \\
    =&
    \delta_{p,q}
    \left(
    \delta_{p,L}
    (-1)^{r-1}(L-r)
    -
    (1-\delta_{p,L})
    \frac{\sin\frac{\pi pr}{L}}{\sin\frac{\pi p}{L}}
    \right)
    -
    (1-\delta_{p,q})
    \delta_{p,q}^{(\mathrm{mod}\, 2)}
    \frac{\sin\frac{\pi(p+q)r}{2L}}{\sin\frac{\pi(p+q)}{2L}},
\end{align}
where
\begin{align}
    \delta_{p,q}^{(\mathrm{mod}\, 2)}
    =
    \begin{cases}
        1 & (\text{if \(p=q\) \(\mathrm{mod}\, 2\)})
        \\
        0 & (\text{otherwise})
    \end{cases}.
\end{align}
Using these results, \(M_{pq}\) is evaluated as follows
\begin{align}
    \label{eq:Mresult}
    M_{pq}
    =&
    \left(
    \sum_{r=1}^{L-1}2(-1)^{r}J_{r}\left(1-\frac{r}{L}\right)\cos\left(\frac{\pi pr}{L}\right)
    \right)
    \delta_{p,q}
    +
    M_{pq}',
\end{align}
where
\begin{align}
    \label{eq:Mprime}
    M_{pq}'
    =&
    \frac{1-\delta_{p,L}}{L}
    \sum_{r=1}^{L-1} 
    2(-1)^{r}J_{r}
    \frac{\sin\frac{\pi pr}{L}}{\sin\frac{\pi p}{L}}
    \delta_{p,q}
    \notag
    \\
    &-
    \sum_{r=1}^{L-1} 
    \frac{(-1)^{r}J_{r}}{L}
    \sqrt{(2-\delta_{p,L})(2-\delta_{q,L})}
    \delta_{p,q}^{(\mathrm{mod}\, 2)}
    (1-\delta_{p,q})
    \left(
    \frac{\cos\frac{\pi(p+q)r}{2L}\sin\frac{\pi(p-q)r}{2L}}{\sin\frac{\pi(p-q)}{2L}}
    -
    \frac{\cos\frac{\pi(p-q)r}{2L}\sin\frac{\pi(p+q)r}{2L}}{\sin\frac{\pi(p+q)}{2L}}
    \right).
\end{align}
If \(M_{pq}'\) can be ignored, Eq.~\eqref{eq:Mresult} is diagonalized and the eigenvalues are given by
\begin{align}
    \mathcal{E}(p)
    =
    \sum_{r=1}^{L-1}2(-1)^{r}J_{r}\left(1-\frac{r}{L}\right)\cos\left(\frac{\pi pr}{L}\right).
\end{align}
The minimum eigenvalue is obtained at \(p = 1\) since \(\mathcal{E}(p)\) increases monotonically within the range \(1 \leq p < L\) as
\begin{align}
    \frac{d}{dp}\mathcal{E}(p)
    =
    \frac{\pi J_0}{L}\tan \frac{\pi p}{2L}
    \left(1+\frac{(-1)^{L}}{L}U_{L-1}\left(\cos \frac{\pi p}{L}\right)\right)
    >0,
\end{align}
where we have used the fact that the Chebyshev polynomial of the second kind, \(U_{L-1}(\cos t) = \sin(Lt)/\sin t\), satisfies \(|U_{L-1}(\cos t)| < L\) for \(-1 < \cos t < 1\). Now we show that \(M_{pq}'\) can be ignored for modes with \(p\ll L\). In this case, Eq.~\eqref{eq:Mprime} suggests that
\begin{align}
    M_{pq}'
    =&
    \delta_{p,q}
    O\left(\frac{\sum_{r=1}^{L-1}(-1)^{r}J_{r}}{L}\right)
    +
    (1-\delta_{p,q})
    O\left(\frac{\sum_{r=1}^{L-1}(-1)^{r}J_{r}}{L^2}\right)
\end{align}
because
\begin{align}
    &\frac{\cos\frac{\pi(p+q)r}{2L}\sin\frac{\pi(p-q)r}{2L}}{\sin\frac{\pi(p-q)}{2L}}
    -
    \frac{\cos\frac{\pi(p-q)r}{2L}\sin\frac{\pi(p+q)r}{2L}}{\sin\frac{\pi(p+q)}{2L}}
    =
    O\left(\frac{1}{L}\right).
\end{align}
Thus, we have
\begin{align}
    \sum_{q=1}^L M_{pq}\tilde{a}_q
    =&
    \mathcal{E}(p)\tilde{a}_p
    +
    \sum_{q=1}^L M_{pq}'\tilde{a}_q
    \notag
    \\
    =&
    \mathcal{E}(p)\tilde{a}_p
    +
    \sum_{q=1}^L 
    \delta_{p,q}
    O\left(\frac{\sum_{r=1}^{L-1}(-1)^{r}J_{r}}{L}\right)
    \tilde{a}_q
    +
    \sum_{q=1}^L
    (1-\delta_{p,q})
    O\left(\frac{\sum_{r=1}^{L-1}(-1)^{r}J_{r}}{L^2}\right)
    \tilde{a}_q
    \notag
    \\
    =&
    \mathcal{E}(p)\tilde{a}_p
    +
    O\left(\frac{\sum_{r=1}^{L-1}(-1)^{r}J_{r}}{L}\right)
    \tilde{a}_p
    +
    O\left(\frac{\sum_{r=1}^{L-1}(-1)^{r}J_{r} \|\tilde{\boldsymbol{a}}\|}{L}\right),
\end{align}
\end{widetext}
which implies that the contribution from \(M_{pq}'\), represented by the second and third terms in the final line, can be neglected for \(L\to \infty\) in the secular equation. This result suggests that, for large \(L\), the eigenvectors corresponding to small eigenvalues are expressed in terms of \(\phi_p(k)\), and in particular, the eigenvector of the smallest eigenvalue is represented by \(\phi_1(k)\) as shown in Eq.~\eqref{eq:akapprox}.

\bibliography{refs}
\end{document}